\documentclass[prd,eqsecnum,twocolumn,showpacs,amsfonts,amssymb]{revtex4}

\usepackage{graphicx}

\usepackage{bm}

\newcommand{\beq}{\begin{equation}}
\newcommand{\eeq}{\end{equation}}
\newcommand{\beqs}{\begin{eqnarray}}
\newcommand{\eeqs}{\end{eqnarray}}

\newcommand{\drawsquare}[2]{\hbox{%
\rule{#2pt}{#1pt}\hskip-#2pt%  left vertical
\rule{#1pt}{#2pt}\hskip-#1pt%  lower horizontal
\rule[#1pt]{#1pt}{#2pt}}\rule[#1pt]{#2pt}{#2pt}\hskip-#2pt%  upper horizontal
\rule{#2pt}{#1pt}}% right vertical
\newcommand{\fund}{\raisebox{-.5pt}{\drawsquare{6.5}{0.4}}}%  fund
\begin{document}

\title{Technicolor Models with Color-Singlet Technifermions and their 
Ultraviolet Extensions} 

\author{Thomas A. Ryttov}
%\thanks{thomas.ryttov@stonybrook.edu}

\author{Robert Shrock}
%\thanks{robert.shrock@stonybrook.edu}

\affiliation{
C. N. Yang Institute for Theoretical Physics \\
Stony Brook University \\
Stony Brook, NY 11794}

\begin{abstract}

We study technicolor models in which all of the technifermions are
color-singlets, focusing on the case in these fermions transform according to
the fundamental representation of the technicolor gauge group. Our analysis
includes a derivation of restrictions on the weak hypercharge assignments for
the technifermions and additional color-singlet, technisinglet fermions 
arising from the necessity of avoiding stable bound states with exotic electric
charges. Precision electroweak constraints on these models are also discussed.
We determine some general properties of extended technicolor theories
containing these technicolor sectors.

\end{abstract}


\maketitle

\section{Introduction}

Electroweak symmetry breaking (EWSB) may occur dynamically, via the formation
of bilinear condensates $\langle \bar F F \rangle$ of a set of fermions $\{ F
\}$ subject to a vectorial, asymptotically free gauge interaction, generically
called technicolor (TC), that becomes strongly coupled at the TeV scale
\cite{tc}.  To communicate this symmetry breaking to the Standard-Model (SM)
fermions, which are technisinglets, one embeds technicolor in a larger,
extended technicolor (ETC) gauge theory \cite{etc}.  This involves gauging the
generational index and combining it with the technicolor index. We denote the
generational, technicolor, and ETC gauge groups as $G_{gen.}$, $G_{TC}$, and
$G_{ETC}$.  It follows that $G_{ETC} \supset G_{gen.} \otimes G_{TC}$.  In one
class of technicolor models, technifermions form a Standard-Model family.  In
these models, the ETC gauge bosons are singlets under the SM gauge group,
$G_{SM}= {\rm SU}(3)_c \otimes G_{EW}$, where $G_{EW} = {\rm SU}(2)_L \otimes
{\rm U}(1)_Y$, and the generators of $G_{ETC}$ commute with those of $G_{SM}$:
$[G_{ETC}, \, G_{SM}] = \emptyset$.  

However, the basic aim of technicolor, to break $G_{EW}$ to electromagnetic
U(1)$_{em}$ dynamically, can be realized using purely color-singlet
technifermions.  A minimal technicolor model of this type includes one
SU(2)$_L$ doublet of left-handed technifermions, together with the
corresponding SU(2)$_L$-singlet right-handed technifermions, all of which are
color-singlets.  To maintain the vectorial nature of the technicolor gauge
symmetry (as is necessary in order that it does not self-break when it forms
condensates), the left- and right-handed chiral components of the
technifermions transform according to the same representation of $G_{TC}$. We
use the abbreviations 1FTC and 1DTC for one-\underline{f}amily and
one-SU(2)$_L$-techni\underline{d}oublet TC models, respectively \cite{1dtc}.
Since ETC gauge bosons mediate transitions that take quarks to technifermions,
it follows that in technicolor models in which the technifermions are
color-singlets, some of these ETC gauge bosons transform according to the
fundamental and conjugate fundamental representations of SU(3)$_c$.  Hence, in
these models, commutators of the associated generators of $G_{ETC}$ transform
as the singlet and adjoint of SU(3)$_c$, $[G_{ETC}, \, G_{SM}] \ne \emptyset$,
and
\beq
G_{ETC} \supset {\rm SU}(3)_c \otimes G_{gen.} \otimes G_{TC} 
\quad {\rm for \ 1DTC} \ . 
\label{getcform}
\eeq

In this paper we shall construct and study technicolor models in which all of
the technifermions are color-singlets and that are minimal, in the sense of
being of 1DTC type.  There are several motivations for this work.  One is that
1DTC models can reduce technicolor corrections to $W$ and $Z$ propagators,
since they involve fewer technifermions than 1FTC models.  Another is that 1FTC
models predict technivector mesons that transform as color octets, and this
prediction is in significant tension with lower bounds on the masses of such
particles obtained by the ATLAS and CMS experiments at the Large Hadron
Collider (LHC), as discussed further below. Yet another motivation is that 1FTC
models have a very large global chiral symmetry, and when this is broken
spontaneously via the formation of bilinear technifermion condensates, there
may be problematically light (pseudo)-Nambu Goldstone bosons.  Recently, there
has much considerable interest in 1DTC models \cite{fs}-\cite{fusa}; reviews
include Refs. \cite{sanrev,discoveringtc}.  Much of this work has made use of
the group SU(2)$_{TC}$ with two technifermions in the adjoint
representation (equivalently, the vector representation of an SO(3)$_{TC}$
group).  Here we will give a general discussion that focuses on 1DTC models
with technifermions in the fundamental representation of the technicolor gauge
group.  Our model-building will focus on an SU(3)$_{TC}$ model, but, as will be
seen, a number of our results, such as restrictions on hypercharge assignments,
will apply rather generally to 1DTC models.

This paper is organized as follows.  In Sect. II, to provide some background,
we review one-family technicolor models and their ultraviolet extension to ETC
theories.  Section III contains preliminaries on 1DTC models and Sect. IV
contains some discussion of properties of models of this type with an
SU(2)$_{TC}$ gauge group and technifermions in the fundamental and adjoint
representations.  In Sect. V we study 1DTC models with an SU(3)$_{TC}$ gauge
group and technifermions in the fundamental representation.  In Sect. VI we
derive some properties of ETC ultraviolet extensions of the SU(3)$_{TC}$
theory.  Section VII contains some concluding remarks.

\section{Background on One-Family TC/ETC Models}

To provide a contrasting background perspective for our study of TC/ETC models
with color-singlet technifermions, we briefly discuss one-family TC/ETC models,
which do contain some color-triplet technifermions.  We take $G_{TC} = {\rm
SU}(N_{TC})_{TC}$ and $G_{gen.} = {\rm SU}(N_{gen.})$, where $N_{gen.}=3$ is
the observed number of SM fermion generations.  In the most compact model of
this type, $G_{ETC}$ contains $G_{gen.}  \otimes G_{TC}$ as a maximal 
subgroup. This is arranged by setting
\beq
G_{ETC} = {\rm SU}(N_{ETC})_{ETC} \ , 
\label{sunetc}
\eeq
where 
\beq
N_{ETC} = N_{gen.}+N_{TC} = 3+N_{TC}  \ . 
\label{netc}
\eeq
In accordance with this, one assumes that the technifermions form one SM
family.  One may assign the SM fermions and technifermions to the following
representations of $G_{SM} \otimes G_{ETC}$
\beqs
Q_L = {u \choose d}_L & : & \ (3,2,N_{ETC})_{1/3,L} \cr\cr
u_R & : & \ (3,1,N_{ETC})_{4/3,R} \cr\cr
d_R & : & \ (3,1,N_{ETC})_{-2/3,R} 
\label{techniquarks_1ftc}
\eeqs
and 
\beqs
L_L = {\nu \choose \ell}_L & : & \ (1,2,N_{ETC})_{-1,L} \cr\cr
\nu_R & : & \ (1,1,N_{ETC})_{0,R} \cr\cr
\ell_R & : & \ (1,1,N_{ETC})_{-2,R} \ . 
\label{technileptons_1ftc}
\eeqs
Here the numbers in parentheses are the dimensions of the representations of
the three non-Abelian factor groups in $G_{SM} \otimes G_{ETC}$, the subscript
denotes the weak hypercharge, and we suppress ETC and color indices.
Indicating these indices explicitly, we have, for example, $u_R = u^{ai}_R$,
where $a=1,2,3$ is the color index and $i=1,...,N_{ETC}$ is the ETC index. The
ETC indices $i$ are ordered such that $i=1,2,3$ are generation indices (with
$u^{a1} = u^a$, $u^{a2} = c^a$, $u^{a3}=t^a$, $\ell^1=e$, $\ell^2=\mu$, etc.)
and $4 \le i \le N_{ETC}$ are TC indices.  It will be useful to distinguish
between generational and technicolor indices, and so if $i$ is in the interval
$4 \le i \le N_{ETC}$, we shall usually label it as $\tau$, standing for TC.

If one makes the minimal choice for $G_{TC}$, namely $G_{TC}={\rm SU}(2)_{TC}$,
then, by Eq.  (\ref{netc}), it follows that $G_{ETC}={\rm SU}(5)_{ETC}$.
Detailed studies with reasonably ultraviolet-complete ETC models of this type
were carried out in Refs. \cite{at94}-\cite{sml}. Refs. \cite{ckm,kt} also
presented a critical assessment of an alternate type of TC/ETC model in which
down-type quarks and charged leptons of opposite chiralities are assigned to
relatively conjugate representations of SU(5)$_{ETC}$, while up-type quarks of
opposite chiralities are assigned to the same representations of this group.

In order to account for the hierarchy in the three generations of SM quarks and
charged leptons, the ETC gauge symmetry should break in a sequence of scales
$\Lambda_{ETC,i}$, where $i=1,2,3$, down to the residual exact technicolor
gauge symmetry.  The studies of Refs. \cite{nt}-\cite{ckm} demonstrated how the
sequential breaking of the ETC gauge symmetry can occur. A typical set of ETC
breaking scales used in these studies is
\beqs
\Lambda_{ETC,1} & \simeq & 10^3 \ {\rm TeV}, \cr\cr
\Lambda_{ETC,2} & \simeq & 10^2 \ {\rm TeV}, \cr\cr
\Lambda_{ETC,3} & \simeq & \ {\rm few \  TeV} 
\label{lametc}
\eeqs
for the three SM generations $i=1,2,3$ \cite{notone}.  Having an explicit and
reasonably ultraviolet-complete ETC theory, it was possible to calculate
flavor-changing neutral current (FCNC) processes, and it was shown that in ETC
theories in which the techniquarks transform in a vectorial manner under
$G_{ETC}$, these can be adequately suppressed because of approximate residual
generational symmetries \cite{nt,ckm,kt}.  A mechanism was also presented for
obtaining lepton mixing and small neutrino masses \cite{nt,lrs,ckm}. The
resultant technicolor gauge interaction confines and breaks chiral symmetry at
the scale $\Lambda_{TC}$, thereby producing electroweak symmetry breaking. The
$W$ and $Z$ pick up masses given to leading order by
\beq
m_W^2 = m_Z^2 \cos^2\theta_W = \frac{g^2 N_D f_{TC}^2}{4} \ , 
\label{mwsq}
\eeq
where $g$ is the SU(2)$_L$ gauge coupling and $N_D$ denotes the number of
SU(2)$_L$ doublets of technifermions.  For one-family TC models, $N_D=4$, so
$f_{TC} \simeq 125$ GeV.

The ETC interactions lead to a mass for a fermion of the $i$'th generation of
the generic form
\beq
m_{f^i} \sim \frac{\kappa \eta \Lambda_{TC}^3}{\Lambda_{ETC,i}^2} \ , 
\label{mfi}
\eeq
where $\kappa$ is a numerical constant of order 10,
\beq
\eta = \exp \Big [ \int_{\Lambda_{TC}}^{\Lambda_w} \frac{d\mu}{\mu} \, 
\gamma(\alpha_{_{TC}}(\mu)) \Big ]
\label{eta}
\eeq
is a renormalization-group factor, $\gamma$ is the mass anomalous dimension of
the technifermions, and $\alpha_{_{TC}}(\mu) = g_{_{TC}}(\mu)^2/(4\pi)$ is the
TC running coupling (inherited from its UV completion in the ETC theory),
depending on the Euclidean reference momentum, $\mu$.  In Eq. (\ref{eta}),
$\Lambda_w$ is the scale where the coupling rises to O(1).  The
$\Lambda_{TC}^3$ factor in Eq. (\ref{mfi}) arises from the technifermion
condensate $\langle \bar F F \rangle$ and the $1/\Lambda_{ETC,i}^2$ factor
arises from the propagator(s) of ETC gauge bosons mediating the transitions
between SM fermions of the $i$'th generation and technifermions.  As is evident
in Eq. (\ref{mfi}), the largest $\Lambda_{ETC,i}$ corresponds to the smallest
fermion masses, namely those of the first generation, and so forth for the
other generations.  It has also been shown how off-diagonal elements can be
generated in the full $3 \times 3$ fermion mass matrices, whose diagonalization
thus leads to quark and lepton mixing \cite{nt,ckm}.  Related work is in 
Ref. \cite{lanemartin}. 

Viable TC/ETC theories require a value of $\gamma$ that is not too small, in
order to enhance SM fermion mass generation via the renormalization-group
factor $\eta$.  This property can follow naturally if the theory has an
approximate infrared fixed point (IRFP), i.e., IR zero of the TC beta function,
so that the running TC coupling $\alpha_{_{TC}}(\mu)$ becomes large at a scale
$\Lambda_w$ but runs slowly (walks) as a function of the scale $\mu$
\cite{chipt}. 

One-family technicolor models are subject to many constraints, such as those
from precision electroweak data, neutral flavor-changing current processes,
etc.  In addition, 1FTC models predict color-octet technihadrons, in
particular, pseudoscalar and vector technimesons.  The vector technimesons are
expected to have masses given by
\beq
\frac{m_{V,TC}}{m_{\rho,\omega}} \simeq  \frac{\Lambda_{TC}}{\Lambda_{QCD}} 
\simeq \frac{f_{TC}}{f_\pi} \Big ( \frac{N_c}{N_{TC}} \Big )^{1/2} \ , 
\label{mvtmv}
\eeq
where $N_c=3$, $f_\pi \simeq 93$ MeV, $\Lambda_{QCD} \simeq 330$ MeV, and
$f_{TC} \simeq 125$ GeV, so $m_{V,TC} \simeq 1.0 \sqrt{3/N_{TC}}$ TeV. This
simple scaling behavior is approximately borne out in more sophisticated
calculations of technimeson masses using solutions of the Bethe-Salpeter
equation in a technicolor theory with walking behavior \cite{sg}.  A similar
scaling is expected to apply for the width $\Gamma_{V,TC}$, so that
$\Gamma_{V,TC}$ would be a few hundred GeV.  The ATLAS and CMS experiments at
the LHC have set a lower limit of approximately 2 TeV on color-octet resonances
of this type \cite{cms_colormesons,atlas_colormesons}. There is thus
significant tension between these LHC data and one-family technicolor models.
This tension is exacerbated by limits on the (pseudo)-Nambu-Goldstone bosons,
denoted (P)NGBs, in this model \cite{ehlq}-\cite{pdg}.  As noted above, this
provides one motivation for exploring TC/ETC models that contain only
color-singlet technifermions, since these technifermions thus do not couple
directly to gluons, and hence the resultant technihadrons are not subject to
such severe experimental limits from current LHC (or Tevatron) data.

\section{Technicolor Models with Color-Singlet Technifermions}

In this section we begin our discussion of 1DTC models. As will be 
explained below, these models may, in general, also contain other
technifermions that are $G_{SM}$-singlets. These models have the feature that
all technifermions are SU(3)$_c$-singlets and can thus be denoted also as
color-singlet technifermion (CSTF) theories.  For the models of
interest here, the 1DTC property implies the CSTF property.  The converse does
not hold in general, since, in principle, a technicolor model could
contain only color-singlet technifermions but have more than one SU(2)$_L$
technidoublet.  However, in practice, given the requirement of minimizing
technicolor corrections to the $W$ and $Z$ propagators, as long as one
considers CSTF models, one restricts to those of 1DTC type. Hence, in practice,
one has the relation 1DTC $\Leftrightarrow$ CSTF for these properties.

The gauge symmetry that is operative at an energy scale of $\sim 1$ TeV is
assumed to be $G_{SM} \otimes G_{TC}$.  We shall mainly focus on the case in
which the technifermions to be in the fundamental representation of $G_{TC} =
{\rm SU}(N_{TC})_{TC}$ (while sometimes giving more general results), and shall
consider the possible choices $G_{TC}={\rm SU}(2)_{TC}$ and $G_{TC}={\rm
SU}(3)_{TC}$.  The technicolor model is minimal in the sense that it uses the
minimum content of $G_{EW}$-nonsinglet technifermions necessary to achieve
electroweak symmetry breaking, with the left-handed and right-handed components
of the technifermions transforming as
\beq
F^\tau_L = {f^\tau_1 \choose f^\tau_2}_L \ : \ (1,2,d_{R_{TC}})_{Y_{F_L}}
\label{fl}
\eeq
and
\beq 
f_{1R}^\tau: \ (1,1,d_{R_{TC}})_{Y_{f_{1R}}} \ , \quad 
f_{2R}^\tau: \ (1,1,d_{R_{TC}})_{Y_{f_{2R}}}
\label{fr}
\eeq
under $G_{SM} \otimes G_{TC}$, where $d_{R_{TC}}$ denotes the dimension of the
representation $R_{TC}$, and we again suppress the TC index $\tau$ in the
notation. The electric charge operator is $Q=T_{3L}+(Y/2)$ (in units of $e$),
so the condition that U(1)$_{em}$ be vectorial on these technifermions means
that
\beq
\frac{1}{2} + \frac{Y_{F_L}}{2} = q_{f_{1L}} = q_{f_{1R}}= 
               \frac{Y_{f_{1R}}}{2}
\label{f1qrel}
\eeq
and
\beq
-\frac{1}{2} + \frac{Y_{F_L}}{2} = q_{f_{2L}} = q_{f_{2R}}= 
\frac{Y_{f_{2R}}}{2} \ . 
\label{f2qrel}
\eeq
Hence, 
\beq
1+Y_{f_L} = Y_{f_{1R}} \ , \quad -1+Y_{f_L} = Y_{f_{2R}}\ .
\label{ylr}
\eeq
The Lagrangian mass terms $m_{f_1}\bar f_{1\tau L} f^\tau_{1R} + h.c.$ and
$m_{f_2} \bar f_{2\tau L} f^\tau_{2R} + h.c.$ would explicitly break ${\rm
SU}(2)_L \otimes {\rm U}(1)_Y$, and therefore these are set to zero.

\section{SU(2)$_{TC}$ 1DTC Models} 

Here we take $G_{TC}={\rm SU}(2)_{TC}$.  If $Y_{F_L}=0$, then the theory with
the fermions (\ref{fl}) and (\ref{fr}) is free of any anomalies in gauged
currents.  It is also free of any global SU(2)$_L$ anomaly, since it contains
an even number, $N_{gen.}(N_c+1)+N_{TC}=14$ chiral SU(2)$_L$ doublets of
fermions.  (Of course, there is also no global SU(2)$_{TC}$ anomaly whether the
number of technifermions transforming as fundamental representations of
SU(2)$_{TC}$ is even or odd, since these are Dirac fermions, corresponding to
the fact that the SU(2)$_{TC}$ theory is vectorial.)  However, this model is
disfavored by the fact that, as $\alpha_{_{TC}}$ grows to a size of order unity
and the TC interaction eventually confines and produces bilinear technifermion
condensates and associated spontaneous chiral symmetry breaking at the scale
$\Lambda_{TC}$, these condensates would most likely have an undesired form. The
condensates form in the most attractive channel (MAC), which is $2 \times 2 \to
1$.  Vacuum alignment arguments imply that the condensates should preserve the
maximal possible gauge symmetry and hence would have the Majorana forms
\beq
\langle \epsilon_{\alpha \beta}\epsilon_{\tau \tau'} F^{\alpha \tau \ T}_L C 
F^{\beta \tau'}_L \rangle = 2\langle \epsilon_{\tau \tau'} f^{\tau \ T}_{1L} C 
f^{\tau'}_{2L} \rangle
\label{flflcondensate}
\eeq
and
\beq
\langle \epsilon_{\tau \tau'} f^{\tau \ T}_{1R} C f^{\tau'}_{2R} \rangle \ , 
\label{frfrcondensate}
\eeq
where here $\alpha,\beta$ are SU(2)$_L$ indices and $\tau,\tau'$ are
SU(2)$_{TC}$ indices.  These condensates are invariant under all gauge
symmetries, in particular, $G_{EW}$.  Hence, they would not achieve the basic
purpose of technicolor, which is electroweak symmetry breaking.

One might try to avoid this by assigning a nonzero weak hypercharge $Y_{F_L}$
to $F_L$, which, by Eq. (\ref{ylr}), would imply that at least one of the
$f_{jR}$, $j=1,2$, would also have nonzero weak hypercharge. This modification
would contain nonzero gauge anomalies unless one also added more fermions. 
A simple approach is to add an even number of (color-singlet)
technicolor-singlet fermions that transform as doublets under SU(2)$_L$.  This
number must be even to avoid producing a global SU(2)$_L$ anomaly.  The minimal
such number is two, and thus we add the following color-singlet,
technicolorsinglet fermions, forming two left-handed SU(2)$_L$ doublets and 
four corresponding right-handed SU(2)$_L$-singlets
\beq
\psi_L = {\psi_1 \choose \psi_2}_L  \ , \quad \psi_{jR} \ , \quad j=1,2, 
\label{psi}
\eeq
and
\beq
\psi'_L = {\psi_1' \choose \psi_2'}_L  \ , \quad \psi_{jR}' \ , \quad j=1,2, 
\label{psiprime}
\eeq
with respective weak hypercharges $Y_{\psi_L}$, $Y_{\psi_{1R}}$,
$Y_{\psi_{2R}}$ and $Y_{\psi'_L}$, $Y_{\psi'_{1R}}$, $Y_{\psi'_{2R}}$.  It is a
model-building choice whether or not one attributes a nonzero lepton number to
these fermions. We do not make a definite commitment concerning this choice and
hence do not refer to the $\psi_i$ or $\psi'_i$ as leptons, but simply as
color-singlet, technicolor-singlet fermions.  The hypercharges $Y_{\psi_L}$,
$Y_{\psi_{1R}}$, $Y_{\psi_{2R}}$ satisfy relations analogous to
(\ref{f1qrel})-(\ref{ylr}),
\beq
\frac{1}{2} + \frac{Y_{\psi_L}}{2} = q_{\psi_{1L}} = q_{\psi_{1R}}=
\frac{Y_{\psi_{1R}}}{2}
\label{psi1rel}
\eeq
and
\beq
-\frac{1}{2} + \frac{Y_{\psi_L}}{2} = q_{\psi_{2L}} = q_{\psi_{2R}}=
\frac{Y_{\psi_{2R}}}{2}
\label{psi2qrel}
\eeq
so that 
\beq
1+Y_{\psi_L} = Y_{\psi_{1R}} \ , \quad -1+Y_{\psi_L} = Y_{\psi_{2R}} \ .
\label{yllr}
\eeq
The $Y_{\psi'_L}$, $Y_{\psi'_{1R}}$, $Y_{\psi'_{2R}}$ satisfy the analogous
relations with primes.  Then the gauge anomalies of the form 
\beq
{\rm U}(1)_Y^3, \quad {\rm SU}(2)_L^2 \, {\rm U}(1)_Y,  \quad 
Gr^2 \, {\rm U}(1)_Y \ , 
\label{anomlist}
\eeq
where $Gr$ denotes graviton, are satisfied if and only if 
\beq
d_{R_{TC}} \, Y_{F_L} + Y_{\psi_L} + Y_{\psi'_L} = 0 \ . 
\label{anomalycancelprime}
\eeq
Explicitly, this equation reads $2Y_F+Y_\psi+Y_{\psi'}=0$. 
The solutions of this condition form a two-parameter set.  In this case, it 
is possible that the condensation of the technifermions might proceed in the
desired manner, yielding the Dirac condensates
\beq
\langle \bar f_{i \tau L} f^\tau_{iR}\rangle \ , \quad i=1,2 \ . 
\label{diraccondensate}
\eeq
These condensates are equal for $i=1$ and $i=2$, up to small corrections from
weak hypercharge interactions, and hence so are the associated dynamical
technifermion masses for $f_1$ and $f_2$.  The condensates
(\ref{diraccondensate}) for each $i=1,2$ break $G_{EW}$ to
U(1)$_{em}$, as desired.  They transform as $\Delta T_{3L}=1/2$ and $\Delta
Y=1$ operators and hence yield the tree-level mass relation $\rho=1$, where 
\beq
\rho \equiv \frac{m_W^2}{m_Z^2 \cos^2\theta_W} \ , 
\label{rho}
\eeq
as is necessary to agree with experiment.  

However, this desired pattern of condensation is not guaranteed.  It is also
possible that, even with nonzero weak hypercharge assignments, the theory would
still prefer to form the Majorana condensates (\ref{flflcondensate}) and
(\ref{frfrcondensate}) instead of the Dirac condensates
(\ref{diraccondensate}). If this happened, then these Majorana condensates
(\ref{flflcondensate}) and (\ref{frfrcondensate}) would break only the U(1)$_Y$
part of $G_{EW}$, while preserving the SU(2)$_L$ part.  Both of the condensates
(\ref{flflcondensate}) and (\ref{frfrcondensate}) transform as $\Delta Y =
2Y_{F_L}$ operators.  Indeed, this latter type of condensation could actually
be favored by a vacuum alignment argument on the grounds that it only breaks
one of the four generators of $G_{EW}$, while the Dirac condensates
(\ref{diraccondensate}) break all of the generators, leaving one linear
combination (the electric charge, $Q$) invariant.  The use of the SU(2)$_{TC}$
gauge group in the one-family technicolor models does not encounter this
problem because in that case the Majorana-type condensates would break
SU(3)$_c$ and U(1)$_{em}$ and hence are excluded by a vacuum alignment argument
\cite{lrsft}. 

 One way of avoiding this problem with undesired condensates that has been
investigated is to assign the technifermions to an adjoint representation of
SU(2)$_{TC}$ \cite{fs}-\cite{discoveringtc}. This is also useful in reducing
technicolor contributions to $W$ and $Z$ propagator corrections.  This model is
often denoted the minimal walking technicolor (MTW) model, since, first, it
includes a minimal content of $G_{EW}$-nonsinglet technifermions, and,
secondly, it can achieve walking behavior with a minimal set of technifermions.
For our general discussion here, it will be useful to remark on some properties
of this model, especially concerning vacuum alignment. Since the adjoint
representation of SU(2) is equivalent to the vector representation of SO(3), we
may denote the left-handed technifermions as
\beq
{\vec F}_L = {{\vec f}_1 \choose {\vec f}_2 }_L \ , 
\label{fladjoint}
\eeq
and the right-handed technifermions as ${\vec f}_{1R}$, and ${\vec f}_{2R}$.
The number of SU(2)$_L$ chiral doublets is odd (equal to
$N_{gen.}(N_c+1)+3=15$), and hence one is led to introduce one such SU(2)$_L$,
namely the color-singlet, technisinglet fermions of Eq. (\ref{psi}).  The
condition of zero gauge anomalies is satisfied if and only if
$3Y_{F_L}+Y_{\psi_L}=0$.

A basic question is whether the SU(2)$_{TC}$ theory with two (massless) Dirac
fermions in the adjoint representation evolves into the infrared in the desired
manner, with confinement and spontaneous chiral symmetry breaking via the
formation of bilinear fermion condensates that break $G_{EW}$ to U(1)$_{em}$,
or whether, instead, it evolves to an infrared conformal phase with no such
S$\chi$SB.  Lattice studies suggest that this theory evolves into the infrared
toward an exact IR fixed point (IRFP) which is at a sufficiently small value of
$\alpha_{_{TC}}$ that there is no S$\chi$SB or formation of any fermion
condensates.  Since one needs such condensates for electroweak symmetry
breaking, one would have to add requisite four-fermion terms to allow S$\chi$SB
at a smaller value of $\alpha_{_{TC}}$ \cite{fusa}.  Presuming such a term is
generated by an ultraviolet completion, one then may examine the condensates
that form.  

The most attractive bilinear fermion condensation channel is $3 \times 3 \to
1$, and there are several condensates that, {\it a priori}, could form in this
channel. Since the scalar product of two vectors of SO(3) is symmetric under
interchange of the vectors, while the weak SU(2)$_L$ contraction with
$\epsilon_{\alpha\beta}$ is antisymmetric, the resultant Majorana condensate
vanishes identically:
\beq
\langle \epsilon_{\alpha\beta} {\vec F}^{\alpha \ T}_L \cdot C 
{\vec F}^\beta_L \rangle = 0 \ . 
\label{flvflvcondensate}
\eeq
Hence, the ${\vec f}_{1L}$ and ${\vec f}_{2L}$ must condense via Dirac
condensates with the corresponding right-handed technifermions. A vacuum
alignment argument leads to the conclusion that these condensates are 
\beq
\langle {\bar {\vec f}}_{1L} \cdot {\vec f}_{1R} \rangle \ , \quad 
\langle {\bar {\vec f}}_{2L} \cdot {\vec f}_{2R} \rangle \ . 
\label{flvfrvcondensates}
\eeq
These transform as $\Delta T_{3L} = 1/2$, $\Delta Y = 1$ operators, breaking
$G_{EW}$ in the desired manner to U(1)$_{em}$. The same vacuum alignment 
argument implies that none of the following condensates form: 
(i) the Dirac condensates 
    $\langle {\bar {\vec f}}_{1L} \cdot {\vec f}_{2R} \rangle$ and 
    $\langle {\bar {\vec f}}_{2L} \cdot {\vec f}_{1R} \rangle$, which 
transform as $\Delta T_{3L}=1/2$ but violate charge, and (ii) the Majorana
condensates 
$\langle {\vec f}^T_{1R} \cdot C {\vec f}_{1R} \rangle$ and 
$\langle {\vec f}^T_{2R} \cdot C {\vec f}_{2R} \rangle$, which preserve
    SU(2)$_L$ but violate charge.  For the case $Y_{F_L} \ne 0$, this argument
also implies that there is no formation of the Majorana condensate 
$\langle {\vec f}^T_{1R} \cdot C {\vec f}_{2R} \rangle$, which would 
preserve SU(2)$_L$ and transform as a $\Delta Y= \Delta Q = 2Y_{F_L}$
operator.  If $Y_{F_L}=0$, this condensate could form, but we shall demonstrate
below that the assignment $Y_L=0$ generically leads to a problem with
unobserved exotically charged matter.  We next proceed to investigate a
different class of technicolor models.

\section{SU(3)$_{TC}$ 1DTC Models}

\subsection{General Construction} 

Here we construct and study models with the technicolor gauge group
SU(3)$_{TC}$ and technifermions transforming according to the fundamental
representation of this group.  These technifermions thus comprise the requisite
special case of Eqs. (\ref{fl}) and (\ref{fr}).  Because the number of
SU(2)$_L$ chiral doublets is then odd, namely $N_{gen}(N_c+1)+N_{TC}=12+3=15$,
it is necessary to add an odd number of additional SU(2)$_L$ doublets to avoid
a global SU(2)$_L$ anomaly.  We choose to add the minimal number, viz., one,
with the color-singlet, technicolor-singlet fermions of Eq. (\ref{psi}).  The
resultant theory is free of all anomalies in gauged currents if and only if
\beq
d_{R_{TC}} \, Y_{F_L}+Y_{\psi_L}=0, \quad i.e., \quad 
3Y_{F_L}+Y_{\psi_L}=0  \ . 
\label{anomalycancel}
\eeq
The solutions of this condition form a one-parameter set \cite{tcguts}.
It will be useful to give a general classification of the types of solutions in
this set. First, there are three discrete special cases. We denote these with
the abbreviations ZY, SMY and RSMY, standing for ``zero $Y_{F_L}$ and
$Y_{\psi_L}$'', ``SM-type $Y$'' and ``reversed-sign SM-type $Y$'' assignments:
\beqs
{\rm ZY}: \quad Y_{F_L} & = & Y_{\psi_L}=0 \quad \Rightarrow \cr\cr
q_{f_1} & = & q_{f_2}+1 = q_{\psi_1} = q_{\psi_2}+1 = \frac{1}{2} \cr\cr
& & 
\label{ysym}
\eeqs
and, with $N_{TC}=3$, 
\beqs
{\rm SMY}: 
& & \quad Y_{F_L} = \frac{1}{3}, \quad Y_{\psi_L}=-1 \quad \Rightarrow \cr\cr
& & q_{f_1} = q_{f_2}+1 = \frac{2}{3}, \cr\cr
& & q_{\psi_1} = q_{\psi_2}+1 = 0
\label{smy}
\eeqs
or 
\beqs
{\rm RSMY}: \quad 
& & Y_{F_L} = -\frac{1}{3}, \quad Y_{\psi_L}=1 \quad \Rightarrow \cr\cr
& & q_{f_1} = q_{f_2}+1 = \frac{1}{3}, \cr\cr
& & q_{\psi_1} = q_{\psi_2}+1 = 0 \ . \cr\cr
& & 
\label{rsmy}
\eeqs
Indicating the charges explicitly, we have, for the SU(2)$_L$ doublets,
\beqs
& & {\rm ZY}:\quad F_L = { f^\tau_1(1/2) \choose f^\tau_2(-1/2)}_L  \ , \quad
             \psi_L = { \psi_1(1/2) \choose \psi_2(-1/2)}_L \ , \cr\cr
& & 
\label{qsym}
\eeqs
\beqs
& & {\rm SMY}:\quad F_L = { f^\tau_1(2/3) \choose f^\tau_2(-1/3)}_L \ , \quad
              \psi_L = { \psi_1(0) \choose \psi_2(-1)}_L \ , \cr\cr
& & 
\label{qsmy}
\eeqs
and
\beqs
& & {\rm RSMY}:\quad F_L = { f^\tau_1(1/3) \choose f^\tau_2(-2/3)}_L \ ,\quad
             \psi_L = { \psi_1(1) \choose \psi_2(0)}_L \ , \cr\cr
& & 
\label{qrsmy}
\eeqs
with corresponding charge assignments for the $f^\tau_{jR}$ and $\psi_{jR}$.
Note that in the SMY case, even though the technifermions $f_1$ and $f_2$ have
the same electric charges as the up-type and down-type quarks, respectively,
they cannot mix with these quarks, since this would violate the exact SU(3)$_c$
color symmetry (as well as the exact SU(3)$_{TC}$ technicolor symmetry).  The
same statement applies for the RSMY case, where $f^c_2$ and $f^c_1$ have the
same electric charges as the up-type and down-type quarks, respectively.  The
SMY assignments coincide with the usual ones in the Standard Model, and the
RSMY assignments are obtained by reversing the signs of the hypercharges in the
SMY case.  Only for the SMY and RSMY cases is one of the $\psi_j$ neutral; for
the SMY and RSMY choices, respectively, this is $\psi_1$ and $\psi_2$, as
indicated in the equations above \cite{eha}.  Note that one can equivalently
describe the RSMY case in terms of charge-conjugated fermions with SMY
hypercharge assignments, viz., $F^c_{\tau,R}$, with $Y=1/3$, $L^c_R$ with
$Y=-1$, etc. In this form, the technifermions would transform as conjugate
fundamental, rather than fundamental, representations of SU(3)$_{TC}$. However,
without loss of generality, we will keep the forms as in Eq.  (\ref{rsmy}).

The others in the continuous one-parameter set of solutions of the anomaly
cancellation condition (\ref{anomalycancel}) form four different classes: 
\beqs
{\rm I}: \ & & Y_{\psi_L} > 1, \ \ {\rm so} \ \
q_{\psi_i} > 0 \quad {\rm for \ both} \ \ i=1 \ \ {\rm and} \ \ i=2 \cr\cr
& & 
\label{ppcase}
\eeqs
\beqs
{\rm II}: \ & & Y_{\psi_L} < -1, \ \ {\rm so} \ \ 
q_{\psi_i} < 0 \quad {\rm for \ both} \ \ i=1 \ \ {\rm and} \ \ i=2 \cr\cr
& & 
\label{mmcase}
\eeqs
\beqs
{\rm III}: \ & & -1 < Y_{\psi_L} < 0, \ \ {\rm so} \ \ 
0 < q_{\psi_1} < \frac{1}{2} \cr\cr
 & & {\rm and} \ \ -1 < q_{\psi_2} < -\frac{1}{2} 
\label{minus1tozerocase} 
\eeqs
and
\beqs
{\rm IV}: \ & & 0 < Y_{\psi_L} < 1, \ \ {\rm so} \ \ 
\frac{1}{2} < q_{\psi_1} < 1 \cr\cr
& & {\rm and} \ \ -\frac{1}{2} < q_{\psi_2} < 0 \ . 
\label{zeroto1case} 
\eeqs
Restrictions on these hypercharge assignments will be given below. 

\subsection{Option of Augmenting the Model for Walking Behavior}

As so far constructed, the SU(3)$_{TC}$ theory has only $N_{tf}=2$ (Dirac)
technifermions, which is well below the range of values of $N_{tf}$ where
continuum and lattice studies indicate that walking behavior occurs.  Since
this walking property is desirable to enhance SM fermion masses (provided that
the associated mass anomalous dimension is, in fact, not too small), one thus
looks for ways to augment the fermion content of the theory so as to produce
walking.  In order for this theory to have walking, one would choose the
content of technifermions to be such as to yield an approximate IR fixed point
at a value $\alpha_{_{TC}} = \alpha_{IR}$ that is slightly greater than the
minimum value for spontaneous chiral symmetry breaking, denoted
$\alpha_{cr}$. The key fact that one can make use of is that although some
technifermions must be nonsinglets under $G_{EW}$ with representations as given
in Eqs. (\ref{fl}) and (\ref{fr}), other technifermins may be $G_{EW}$-singlets
and, indeed, fully $G_{SM}$-singlets. Since technifermions transforming like
those in Eqs. (\ref{fl}) and (\ref{fr}) contribute to $W$ and $Z$ propagator
corrections, and since one would like to minimize these contributions, one is
thus naturally led to choose any additional technifermions to be
$G_{EW}$-singlets.  These must also be color-singlets, since otherwise some
technivector mesons would transform as color-octets, and one would encounter
the same problem with the LHC lower limits on the masses of such particles that
one does with one-family technicolor models.  So the additional technifermions
that would be added for walking behavior should be $G_{SM}$-singlets. This type
of device has been used before to get walking, e.g., in \cite{ts}.  Although it
is not mandatory to take these $G_{SM}$-singlet technifermions to transform
according to the fundamental representation, we shall do so here because this
makes possible a simpler embedding of the technicolor model in the eveloping
extended technicolor theory.  As we shall show in a later section, the
structure of the ETC ultraviolet extension of the model is strongly affected by
whether one includes or does not include these additional $G_{SM}$-singlet,
technisinglet fermions.

In contrast to 1FTC models, in which all technifermions are
$G_{SM}$-nonsinglets, in the present type of model some technifermions are not
just color-singlets, but also electroweak singlets.  This fact, together with
the fact that the technicolor gauge interaction is vectorial, means that, at
the technicolor level, no gauge symmetry forbids these $G_{SM}$-singlet
technifermions from having nonzero Lagrangian masses in the effective
Lagrangian that is operative slightly above TeV scale.  Consequently, there
are, in principle, two parameters that we may choose in determining the
structure of the $G_{SM}$-singlet technifermion sector of the augmented model,
namely the overall number of such fermions, and their masses.  We begin with a
discussion of the case in which all of the technifermions have zero Lagrangian
masses and then proceed to remark on the more general case in which some
$G_{SM}$-singlet technifermions have nonzero Lagrangian masses. Throughout this
discussion, it should be recalled that all technifermions gain dynamical masses
of order $\Lambda_{TC}$ from the confinement and formation of
chiral-symmetry-breaking bilinear condensates that form in the technicolor
theory.

Analyses of the ladder approximation to the Dyson-Schwinger equation for the
technifermion propagator suggest that $\alpha_{cr} = \pi/(3 C_f)$, where $C_f$
is the quadratic Casimir invariant for the technifermion representation $R$ of
${\rm SU}(N_{TC})$ \cite{chipt}.  Setting this equal to the two-loop value of
$\alpha_{IR}$ yields an estimate for $N_{tf,cr}$, defined as the value of
$N_{tf}$ such that, for $N_{tf} < N_{tf,cr}$ and $N_{tf} > N_{tf,cr}$, the
theory evolves into the IR with S$\chi$SB and without S$\chi$SB, respectively.
In the former case, where S$\chi$SB occurs, the IRFP is approximate, since when
the technifermions condense and gain dynamical masses, one integrates them out
in the effective low energy field theory applicable below the confinement and
condensation scale $\Lambda_{TC}$, so that the beta function changes to a pure
gauge beta function, which has no perturbative IRFP \cite{nonpert_irfp}.  In
contrast, in the latter case of no S$\chi$SB, the IRFP is exact, and the theory
is conformal in the IR.  For $N_{TC}=3$, this method yields the estimate
$N_{tf,cr} \simeq 12$ \cite{chipt}.  Although the standard analysis of the
Dyson-Schwinger equation neglects instantons, which enhance spontaneous chiral
symmetry breaking, it may still achieve reasonable accuracy because it also
neglects an effect (having to do with a reduction in the integration interval
over virtual Euclidean momenta due to confinement of the technifermions) that
goes in the opposite direction, reducing the tendency to S$\chi$SB \cite{lmax}.
There have been a number of recent lattice studies of this theory
\cite{su3nfc}. Higher-order calculations, up to four-loop order, of the IR zero
of the beta function, i.e., the value of the approximate or exact IRFP, and of
the technifermion mass anomalous dimension $\gamma$ evaluated at this zero,
have also been given \cite{bvh,ps}. Since the model, as constructed so far, has
$N_{wk}=2$ Dirac technifermions in Eqs. (\ref{fl}) and (\ref{fr}), one would
envision adding $N_{tf,cr}-2$ additional massless Dirac technifermions. One
would choose these additional technifermions to be singlets under $G_{EW}$ (as
well as SU(3)$_c$) to ensure that they do not contribute to modifications of
the $W$ and $Z$ propagators. The number of SU(2)$_L$-doublet technifermions is
denoted $N_{tf,ewd}$ and the additional, $G_{EW}$-singlet ($ews$),
technifermions is denoted as $N_{tf,ews}$, so that the total number of
technifermions in the theory is $N_{tf} = N_{tf,ewd}+N_{tf,ews} =
2+N_{tf,ews}$.

More generally, one could allow the possibility that some of the
$G_{SM}$-singlet technifermions may have nonzero Lagrangian masses.  A
constraint on these masses is that they should be small enough, relative to the
scale $\Lambda_w$ where $\alpha_{_{TC}}$ grows to O(1), so that the
technifermions still contribute enough to the beta function coefficients to
give rise to the approximate IRFP that, in turn, yields walking behavior.  For
if this condition were not met, i.e., if some technifermion masses were larger
than $\Lambda_w$, then they would have been integrated out of the low-energy
effective theory applicable at scales below $\Lambda_w$, and thus would not
contribute to the beta function in this theory.  In the absence of all
Lagrangian masses for the technifermions, if one turned off all other gauge
interactions, this theory would have a large (nonanomalous) global chiral
symmetry group ${\rm SU}(N_{tf})_L \otimes {\rm SU}(N_{tf})_R \otimes {\rm
U}(1)_V$, which would be spontaneously broken by the various condensates,
giving rise to various $G_{SM}$-singlet Nambu-Goldstone bosons. Since NGBs have
derivative couplings, their interaction amplitudes are attenuated at low
energies $\sqrt{s}$ by factors of $\sqrt{s}/f_{TC} \propto
\sqrt{s}/\Lambda_{TC}$.  The presence of nonnegligible and, in general,
nondegenerate, Lagrangian masses for these electroweak-singlet technifermions
would reduce the formal global chiral symmetry group and increase the masses of
the (P)NGBs.

\subsection{Instanton Breaking of Number Symmetries}

We next discuss the SU(2)$_L$ instanton-induced breaking of certain global
number symmetries.  By an extension of the analysis carried out in
Ref. \cite{thooft76}, we see that, in addition to the breaking of quark number
$N_q$ and baryon number, $B=N_cN_q=3N_q$ by SU(2)$_L$ instantons, these also
break the number symmetry associated with the $\psi_L$ fields and the
SU(2)$_L$-doublet technifermions, and hence also total technibaryon number,
$B_{TC}$, even though a subset of the fermions contributing to this are
SU(2)$_L$-singlets.  At temperatures low compared with the electroweak scale,
these SU(2)$_L$ instantons are exponentially suppressed, but at temperatures
higher than this scale, they are not suppressed \cite{krs}.

\subsection{Phenomenology of and Constraints on $\psi$ Fermions}

The color-singlet, technisinglet $\psi$ fermions are a notable feature of this
type of TC/ETC model, because if one retains the normal property of the SM,
that quarks and leptons come in families for each generation, then one cannot
assign a generation index to these $\psi$ fermions, since there is no
corresponding fourth generation of quarks.  Note that, even before dealing with
phenomenological constraints on a fourth generation of quarks, one would not
want to add them to this model, since this reinstate the problem of an odd
number of SU(2)$_L$ doublets.  Thus, with the $\psi$ fermions, one has a
qualitatively new kind of (technisinglet) fermion, namely one with no usual
generational index.  Alternately, if one were to consider the $\psi$ fermions
as a fourth generation of leptons, then this would, {\it ipso facto}, redefine
the meaning of the term ``generation'', which hitherto had meant a family of SM
quarks and leptons.  This has important implications for an ETC theory, since
it requires one to postulate a new kind of ETC gauge-mediated transition
between the $\psi$ fermions and technifermions that does not involve the usual
generational index. This transition is necessary in order for ETC interactions
to give masses to these $\psi$ fermions.  Indeed, the masses of $\psi_1$ and
$\psi_2$ must be quite large in order not to conflict with current lower mass
limits from LEP and hadron colliders. (Actually, $\psi_1$ and $\psi_2$ are
group eigenstates and not, in general, mass eigenstates; here and below, when
we refer to the masses of $\psi_j$, $j=1,2$, we mean the primary mass
eigenstates in these group eigenstates.)

The phenomenology of the $\psi$ fermions depends on the hypercharge assignment
that is made to define the model.  We proceed to derive constraints on this
assignment. For the hypercharge assignments ZY and I-IV, neither $\psi_1$ nor
$\psi_2$ is electrically neutral.  In these cases, no mixing can occur between
$\psi_1$ or $\psi_2$ (or their conjugates) and the SM leptons, and, as a
consequence, the lighter member of the set $\{\psi_1, \psi_2 \}$ is stable.
We denote this lighter ($\ell$) member as $\psi_{\ell}$ and the heavier ($h$)
member as $\psi_h$.  In the early universe, as the temperature $T$ decreases
below the scale of the mass $m_{\psi_\ell}$, there will generically be residual
$\psi_\ell$'s or their charge conjugates, $\psi_\ell^c$'s, depending on initial
$\psi$-number asymmetries and physics in the UV completion of the theory that
could give rise to such asymmetries.  First, let us consider the case in which
there is a residual population of $\psi_\ell$ fermions.  There are then two
subcases to analyze.  With hypercharge assignments for which the $\psi_\ell$ is
negatively charged, as the temperature cools sufficiently, this fermion will
form Coulombic bound states with protons, $(p \psi_\ell)$.  With hypercharge
assignments for which the $\psi_\ell$ is positively charged, this fermion will
form Coulombic bound states with electrons, $(e \psi_\ell)$.  We treat these
these subcases in sequence. For the subcase with $q_{\psi_\ell} < 0$, which
leads to a $(p \psi_\ell)$ bound state, the binding energy in the ground state
is, to lowest order in $\alpha_{em}$, \cite{coul},
\beq
E_C[(p \psi_\ell)] = \frac{q_{\psi_\ell}^2 \alpha_{em}^2 \mu_{p\psi_\ell}}{2}
\simeq \frac{q_{\psi_\ell}^2 \alpha_{em}^2 m_p}{2} \ , 
\label{ecoulomb1}
\eeq
where $\mu_{ij}$ is the reduced mass
\beq
\mu_{ij} = \frac{m_i m_j}{m_i + m_j} \ . 
\label{muc1}
\eeq
To infer the last equality of Eq. (\ref{ecoulomb1}), we have used the fact that
$\mu_{p\psi_\ell} \simeq m_p$, since $m_{\psi_\ell} >> m_p$ as required by
current data.  Hence, numerically,
\beq
E_C[(p \psi_\ell)] = (25.0 \ {\rm keV}) \,  q_{\psi_{\ell}}^2 \ . 
\label{ecnumerical1}
\eeq
Similarly, for the subcase with $q_{\psi_\ell} > 0$, which leads to a $(e
\psi_{\ell})$ bound state, the binding energy in the ground state is 
\beq 
E_C[(e \psi_{\ell})]=\frac{q_{\psi_\ell}^2 \alpha_{em}^2 \mu_{e\psi_\ell}}{2}
 \simeq \frac{q_{\psi_\ell}^2 \alpha_{em}^2 m_e}{2} \ ,
\label{ecoulomb2}
\eeq
where here $\mu_{e\psi_\ell} \simeq m_e$, since $m_{\psi_\ell} >> m_e$. 
Hence, numerically, for this subcase, 
\beq
E_C[{e\psi_\ell})] = (13.6 \ {\rm eV}) \,  q_{\psi_{\ell}}^2 \ . 
\label{ecnumerical2}
\eeq
These Coulombic bound states would thus form as $k_BT$ decreases below the
respective $E_C$ values in Eq. (\ref{ecnumerical1}) or (\ref{ecnumerical2}).
They would be stable heavy states with masses in excess of 100 GeV and
non-integral electric charges.  There could also be Coulombic bound states
involving multiple $\psi_{\ell}$'s with higher-$Z$ nuclei in the case where
$q_{\psi_\ell} < 0$. Extensive searches for massive states with exotic,
non-integral charges have been carried out in matter (often as part of free
quark searches), reaching very stringent upper limits on their concentration,
measured in terms of the number fraction $N_{ec}/N_{nuc.}$, where $ec$ 
denotes exotic charge, and $N_{ec}$ and $N_{nuc.}$ are the respective 
numbers of exotic-charge particles and nucleons in a given sample \cite{pdg}
\cite{perl2000}-\cite{perl2009}. Let us denote $Q_{ec}$ as the charge of the
exotically charged ($ec$) particle (in units of $e$, as usual).  These 95 \% CL
upper bounds include \cite{perl2002}
\beq
\frac{N_{ec}}{N_{nucl}} < 1.17 \times 10^{-22} \quad {\rm for} \ \ 
0.18 < |q_{ec}| < 0.82 \ , 
\label{perl2002limit}
\eeq
and \cite{perl2000} 
\beq
\frac{N_{ec}}{N_{nucl}} < 4.71 \times 10^{-22} \quad {\rm for} \ \ 
|q_{ec}|>0.16 \ . 
\label{perl2000limit}
\eeq
Many experiments looking for particles with exotic electric charges have been
motivated by the search for free quarks, and hence have focused on the values
$|q|=1/3$ and $|q|=2/3$ (in units of $e$).  Some of these have reported
considerably more stringent upper limits on $N_{ec}/N_{nucl}$, extending down
to $\sim 10^{-26}$ \cite{homer}. A remark is in order here concerning the
possibility that the hypercharge assignments are such that $|q_{\psi_\ell}| <<
1$.  There have been a number of searches for such electrically charged
particles with charges whose magnitude is much smaller than 1 (in units of
$e$), often called ``milli-charged'' particles \cite{millichargedrev}.  These
have again set very stringent upper limits on such particles.  Recent reviews
of searches for fractionally charged particles include Refs. \cite{pdg} and
\cite{perl2009}.

To complete our discussion, we consider the other case, where the $\psi$-number
asymmetry is such that there is a residual abundance of $\psi^c_\ell$ rather
than $\psi_\ell$ fermions.  Then a similar argument applies. With hypercharge
assignments for which the $\psi^c_\ell$ is negatively charged, as the
temperature cools sufficiently, this fermion will form Coulombic bound states
with protons, $(p \psi^c_\ell)$, and for hypercharge assignments for which the
$\psi^c_\ell$ is positively charged, this fermion will form Coulombic bound
states with electrons, $(e \psi^c_\ell)$.  As before, these are ruled out down
to extremely low number densities by experimental searches. These limits
disfavor hypercharge assignments for which neither $\psi_1$ nor $\psi_2$ is
electrically neutral. 

Combining these results, we infer that the hypercharge assignments ZY and I-IV
are generically disfavored by upper limits on exotic-charged particles in
matter.  This leaves the discrete hypercharge assignments (\ref{smy}) and
(\ref{rsmy}) as being generically allowed.  We note some caveats concerning
this exclusion result. First, it is, in principle, possible that, in contrast
with normal matter, there was a negligibly small $\psi_\ell$-particle number
asymmetry in the early universe and the $\psi_\ell$ and $\psi^c_\ell$ particles
annihilated to very high precision (before forming Coulombic bound states),
leaving an undetectably small residual population of these fermions or
antifermions.  A second type of exception would hold for values of hypercharge
such that the magnitude of the charge $|q_{\psi_\ell}|$ is extremely close to 1
and $m_{\psi_\ell}$ happens to be such that the Coulombic bound states could be
experimentally indistinguishable from neutral atoms of a usual heavy nucleus.
However, we also note that the type of reasoning that we have used to disfavor
various hypercharge assignments is evidently more general than the particular
case of $N_{TC}=3$ and technifermions in the fundamental representation, and
can also be applied to other TC/ETC models \cite{rsf}.  A comment is in order
concerning possible Coulombic bound states of the $\psi$ fermions with
technibaryons.  As will be discussed below, the lightest technibaryon is likely
to be electrically neutral.  Hence, it is unlikely that such bound states would
form.

We discuss some further phenomenology pertaining to the two allowed hypercharge
cases denoted SMY and RSMY.  For the SMY case, the $\psi_1$ is electrically
neutral and would contribute to the invisible decay width of the $Z$ unless its
mass is greater than $m_Z/2$, and similarly for the $\psi_2$ in the RSMY case.
The measurement of the invisible width of the $Z$ by LEP I and its consistency
with three light SU(2)$_L$-doublet neutrinos thus implies that in these two
respective cases, $m_{\psi_\ell} > m_Z/2$, so that these decays are
kinematically forbidden.  Even stronger lower bounds have been obtained from
analyses of LEP II data, which imply that such a fourth SU(2)$_L$-doublet
lepton-like fermion, either neutral or charged, must have a mass greater than
about 90-110 GeV, where the range reflects model-dependent details of how these
mix with, and couple to, the known leptons \cite{pdg,achard01}.

To give the $\psi_i$ masses that are large enough to satisfy these
experimental constraints poses a challenge for this type of model.  With such
large masses, one must also be careful that the model yields a value of the
ratio $m_{\psi_1}/m_{\psi_2}$ that is sufficiently close to unity to avoid an
excessively large contribution to the $\rho$ parameter measuring the violation
of custodial symmetry.  We discuss this further below.

For the SMY case, $\psi_1$ and $\psi_2$ can mix with the usual three
generations of neutrinos and charged leptons, respectively.  These mixings
affect the weak charged-current decays of the $\psi_j$, $j=1,2$.  Moreover,
there will also be decays of the $\psi_j$, $j=1,2$, that are mediated by weak
neutral currents.  To see why these occur, we recall that the necessary and
sufficient conditions for the diagonality of the leptonic neutral weak current
are that leptons of a given electric charge and chirality must have the same
weak $T$ and $T_3$ (equivalently, weak $T$ and $Y$) \cite{ls}.  In general, the
presence of electroweak-singlet neutrinos renders the leptonic neutral weak
current nondiagonal, and hence this is true for the present model, both because
of the $\nu^i_R$ in the Standard Model augmented to include neutrino masses and
because of the $\psi_{1R}$ and $\psi_{2R}$. This follows because we can write a
$\nu_R$ as $\nu^c_L$ for any generation, and $\psi_{jR}$ as
$\psi^c_{jL}$. Consequently, in addition to charged-current decays, there are
also has neutral-current decays of these leptons.  A similar discussion applies
for the RSMY case, where charge conservation allows the $\psi_2$ to mix with
neutrinos and the $\psi_1$ to mix with charged leptons.

\subsection{Some Phenomenology of the Technihadrons} 

Another topic of interest is the technihadrons in the model.  We begin with the
technibaryons.  There are several possibilities here.  As noted above, the
$f_i$ have zero Lagrangian masses, since these would break the electroweak
gauge symmetry.  Because of the formation of the technifermion condensates
(\ref{diraccondensate}) at $\Lambda_{TC}$, these technifermions $f_1$ and $f_2$
gain dynamical masses that are, up to small hypercharge corrections, equal to
each other.  There are several resultant spin-1/2 and spin-3/2 technibaryons
for this $N_{TC}=3$ case.  The classification of these is similar to the
classification of the usual baryons composed of $u$ and $d$ quarks.  The
lightest technibaryons would be the spin-1/2 techninucleons (using lower
indices),
\beq
p_{TC} = (f_1 f_1 f_2) \ , \quad n_{TC} = (f_2 f_2 f_1) \ . 
\label{techninucleons}
\eeq
These have electric charges
\beqs
q_{p_{TC}} & = & 1, \quad  q_{n_{TC}} = 0 \quad {\rm for \ SMY \ case} \cr\cr
q_{p_{TC}} & = & 0, \quad  q_{n_{TC}}= -1 \quad {\rm for \ RSMY \ case}
\label{q_techninucleons}
\eeqs
Since the technifermions have zero Lagrangian masses, and gain dynamical masses
that are equal (of order $\Lambda_{TC}$), up to small electromagnetic
corrections, the techniproton and technineutron are almost degenerate, with
masses given, to leading order, by
\beq
\frac{m_{p_{TC},n_{TC}}}{m_{p,n}} \simeq \frac{\Lambda_{TC}}{\Lambda_{QCD}} 
\simeq \frac{f_{TC}}{f_\pi} \Big ( \frac{N_c}{N_{TC}} \Big )^{1/2} \ . 
\label{mpratio}
\eeq
Hence, with $f_{TC} \simeq 125$ GeV, and $N_{TC}=3$, it follows that 
\beq
m_{p_{TC},n_{TC}} \simeq 1.25 \ {\rm TeV} \ . 
\label{techninucleon_mass}
\eeq
There would be an electromagnetic mass splitting between these techninucleons 
(TCNs) of order 

\beq
|m_{p_{TC}}-m_{n_{TC}}| \sim \frac{\alpha_{em}}{R_{TCN}} \sim
\alpha_{em} \Lambda_{TC} \sim {\rm few \ GeV} \ , 
\label{mtechnipn_massdiff}
\eeq
where $R_{TCN}$ is the spatial size of a techninucleon.  The techninucleon that
is charged (viz., $p_{TC}$ for the SMY case and $n_{TC}$ for the RSMY case) is
heavier, because of its Coulombic self-energy \cite{ems}. For the SMY case, the
$p_{TC}$ would thus decay via a weak charged-current transition to the $n_{TC}$
via the channels
\beqs
{\rm SMY}: \quad p_{TC} & \to & n_{TC} + e^+ + \nu_e, \cr\cr
p_{TC} & \to & n_{TC} + \mu^+ + \nu_\mu, \cr\cr
p_{TC} & \to & n_{TC} + \{ {\rm hadrons} \}^+ \ , 
\label{techniprotondecay_smy}
\eeqs
where in the last line, $\{ \rm hadrons \}^+$ refers to the possible hadronic
final states that can be produced with a few GeV of energy, including $\pi^+$,
$\pi^+\pi^0$, $\rho^+$, and states with higher pion multiplicity.  If the mass
splitting between $p_{TC}$ and $n_{TC}$ is large enough, the decay $p_{TC} \to
n_{TC} + \tau^+ + \nu_\tau$ might also occur, although it would, in any case,
be suppressed by the small phase-space available.  Since both the masses and
the magnitude of the mass splitting for these techninucleons are larger than
those of the actual nucleons by the factor $f_{TC}/f_\pi$, a rough estimate of
the decay rate $\Gamma(p_{TC} \to n_{TC} + \ell^+ + \nu_\ell)$ for $\ell=e$
or $\ell=\mu$ could be obtained by simple scaling as
\beqs
& & \Gamma(p_{TC} \to n_{TC} + \ell^+ + \nu_\ell) \simeq \cr\cr
& & 
\Gamma(n \to p + e^- + \bar\nu_e) \Big ( \frac{\Lambda_{TC}}{\Lambda_{QCD}} 
\Big )^5 \ ,
\label{gamma_ptc_to_leptons}
\eeqs
where $\Gamma(n \to p + e^- + \bar\nu_e) = 1/\tau_n$, with $\tau_n = 0.886
\times 10^3$ sec. and $\Lambda_{TC}/\Lambda_{QCD} \simeq 
(f_{TC}/f_\pi)\sqrt{N_c/N_{TC}} = f_{TC}/f_\pi$ in the present model with
$N_{TC}=3$.  For the inclusive weak decay rate, neglecting phase-space
suppressed modes, we can estimate 
$\Gamma(p_{TC}) = (2+N_c)\Gamma(p_{TC} \to n_{TC} + \ell^+ +
\nu_\ell)$. Combining this with Eq. (\ref{gamma_ptc_to_leptons}), we obtain 
the estimate for the lifetime 
\beq
\tau_{p_{TC}} \sim \frac{1}{5} \, \Big ( \frac{f_\pi}{f_{TC}} \Big )^5 \, 
\tau_n \simeq 10^{-15} \ \ {\rm sec.} 
\label{tauptc}
\eeq
The time $t$ in the early universe by which the temperature $T$ has decreased
to $T \sim m_{p_{TC}} \sim 1$ TeV is $t \sim 10^{-12}$ sec.  After this time,
within a few e-foldings of the lifetime $\tau_{p_{TC}}$, most of the $p_{TC}$
techninucleons would have decayed to $n_{TC}$s's.

An analogous discussion, with obvious changes, applies for the case of RSMY
hypercharge assignments.  Thus, here, 
\beqs
{\rm RSMY}: \quad n_{TC} & \to & p_{TC} + e + \bar\nu_e, \cr\cr
n_{TC} & \to & p_{TC} + \mu + \bar\nu_\mu, \cr\cr
n_{TC} & \to & p_{TC} + \{ {\rm hadrons} \}^- \ . 
\label{techniprotondecay_rsmy}
\eeqs
Similarly, the $n_{TC}$ lifetime, $\tau_{n_{TC}}$, for the RSMY case, would be
essentially equal to $\tau_{p_{TC}}$ for the SMY case.  In each of these two
respective cases, the lightest technibaryon would be stable against weak decay.
Although this technibaryon would be electrically neutral,
it would not be a weakly interacting massive particle (WIMP), for several
reasons.  First, it is composed of electrically charged technifermions, and
these could interact with a photon via magnetic and electric form factors, just
as is the case with the actual neutron.  Second, it would have residual strong
interactions via exchange of technipions (the longitudinal components of $W$
and $Z$ bosons), on the length scale $\sim 1/m_{W,Z}$ and further strong
residual interactions via exchange of technivector mesons, on the length scale
of $\sim 1/(1 \ { \rm TeV})$.  Some related work on technibaryons as possible
sources of dark matter is in Refs.
\cite{nussinov},\cite{mtw_darkmatter},\cite{odense_dm}.  We shall not pursue
this topic here, but it merits further study.

There would also be heavier, spin-3/2 technibaryons, split in mass from these
techninucleons by the technigluon hyperfine interaction, 
\beq
(f_1 f_1 f_1), \ \ (f_1,f_1,f_2), \ \ (f_1,f_2,f_2), \ \ (f_2,f_2,f_2) 
\label{tcdelta}
\eeq
with respective charges $2, \ 1, \ 0, \ -1$ and $1, \ 0, \ -1, \ -2$ for the
SMY and RSMY cases.  The spectrum of the technicolor theory would also contain
mesons with various $J^{PC}$ values and techniglueballs. The three true
Goldstone bosons are absorbed by the $W^\pm$ and $Z$, but there would be some
(pseudo) Nambu-Goldstone bosons (PNGBs) involving the additional
$G_{SM}$-singlet technifermions included for walking behavior.  The masses of
these PNGBs would depend on the bare masses that we assigned to these
additional $G_{SM}$-singlet technifermions.  If these masses were small
compared with $\Lambda_{TC}$, so that the PNGBs were close to being true NGBs,
they would be characterized by derivative couplings and hence would tend to 
decouple at energies low compared with $\Lambda_{TC}$.

\subsection{TC Corrections to $W$ and $Z$ Propagators}

Although perturbation theory cannot be used to estimate the technicolor
contribution to $W$ and $Z$ propagator corrections, since the technicolor
interaction is strongly coupled at the mass scale $m_W$ and $m_Z$, one
nevertheless often refers to the perturbative estimate as a crude guide.  The
most important of these $W$ and $Z$ propagator corrections are embodied in the
$S$ and $T$ parameters \cite{pt}.  As background, we recall that the fermions
in Eq. (\ref{fl}) and (\ref{fr}) have zero Lagrangian masses and gain dynamical
masses from the confinement and spontaneous chiral symmetry breaking at the
scale $\Lambda_{TC}$ due to their technicolor gauge interactions.  Since
$G_{EW}$ interactions are quite weak at the scale $\Lambda_{TC}$, these
dynamical masses of $f_1$ and $f_2$ are equal, up to small corrections.  In
QCD, the constituent (dynamical) quark masses are roughly 330 MeV, while
$f_\pi=93$ MeV.  If one takes the ratio of the dynamical technifermion mass
divided by $f_\pi$ to be roughly similar to the ratio $(330 \ {\rm MeV})/(93 \
{\rm MeV}) = 3.5$ in QCD, then the technifermion dynamical mass $\Sigma_{TC}
\simeq 850$ GeV. Hence, $m_Z^2/\Sigma_{TC}^2 \simeq 1 \times 10^{-2}$.

We recall that for the case with one doublet of fermions, as in Eq. (\ref{fl})
and (\ref{fr}) with masses that are approximately degenerate and are large
compared with $m_Z$, the perturbative contribution to the $S$ parameter is
$\Delta S_{pert.}  = 1/(6\pi)$.  If the technicolor theory has $N_D$ SU(2)$_L$
doublets of technifermions and these transform as the representations $R$ of
$G_{TC}$, then one has
\beq
(\Delta S)_{pert.} = \frac{N_D \, d_{R_{TC}}}{6\pi} \ . 
\label{deltas}
\eeq
For the SU(3)$_{TC}$ theory, $d_{R_{TC}}=3$, so that the perturbative estimate
of the contributions of the technifermions to $S$ is 
$(\Delta S)_{pert.} = 1/(2\pi) \simeq 0.16$. 

We also need to analyze the effects of the $\psi$ fields.  Because these are
technisinglets, they are weakly interacting at the scale $m_Z$, so that their
contributions to loop corrections to the $W$ and $Z$ propagators can be
reliably calculated perturbatively. For the same reason, higher-loop 
corrections due to these $\psi$ fields are expected to be reasonably small 
compared to the one-loop correction.  We use the exact 
one-loop expression for the contribution to $S$, which is (e.g., \cite{hps})
\beqs
(\Delta S)_\psi & = & \frac{1}{6\pi} \Big [ 2(3+2Y_{\psi_L})r_1 + 
2(3-2Y_{\psi_L})r_2 \cr\cr
 & - & Y_{\psi_L}\ln (r_1/r_2) \cr\cr
    & + & \frac{1}{2}\Big [ (3+2Y_{\psi_L})r_1+Y_{\psi_L} \Big ] G(r_1) \cr\cr
    & + & \frac{1}{2}\Big [ (3-2Y_{\psi_L})r_2-Y_{\psi_L} \Big ] G(r_2)\Big ]
 \ , 
\label{spsi}
\eeqs
where 
\beq
r_i = \Big ( \frac{m_{\psi_i}}{m_Z} \Big )^2  
\label{ri}
\eeq
and
\beq
G(r) = -4\sqrt{4r-1} \ {\rm arctan} \Big [ \frac{1}{\sqrt{4r-1}} \Big ] \ . 
\label{g}
\eeq
Note that
\beqs
& & (\Delta S)_\psi \ {\rm is \ invariant \ under \ the \ interchange} \cr\cr
& & m_{\psi_1} \leftrightarrow m_{\psi_2} \quad {\rm with} \ Y_{\psi_L} 
\to -Y_{\psi_L} \ . 
\label{sinvariance} 
\eeqs
It follows that $(\Delta S)_\psi$ does not depend on $Y_{\psi_L}$ if
$m_{\psi_1} = m_{\psi_2}$.  For the experimentally allowed range of
$m_{\psi_j}$, $j=1,2$, lying about 100 GeV, the exact expression is well
approximated by
\beqs
(\Delta S)_{pert.} & = & \frac{1}{6\pi} \bigg [ 
1-2Y_{\psi_L}\ln \Big ( \frac{m_{\psi_1}}{m_{\psi_2}} \Big ) \cr\cr
& + &
 \frac{(1+4Y_{\psi_L})}{20} \Big ( \frac{m_Z}{m_{\psi_1}} \Big )^2 + 
 \frac{(1-4Y_{\psi_L})}{20} \Big ( \frac{m_Z}{m_{\psi_2}} \Big )^2 \cr\cr
& + & O \Big (  \frac{m_Z^4}{m_{\psi_i}^4} \Big ) \bigg ] \ . 
\label{spsi_approx}
\eeqs
If $\psi_1$ and $\psi_2$ are nearly degenerate, they simply contribute an
additional amount $1/(6\pi)$ to $S$, so that, combining this with the rough
perturbative estimate of the technifermion contributions, one would obtain, as
the estimate of the total new addition to $S$, the result $(\Delta S)_{pert}. =
(3+1)/(6\pi) = 2/(3\pi) = 0.21$. Mixing of the $W$ and $Z$ with charged and
neutral technivector mesons, respectively, also affects these corrections.  The
property of walking might reduce this contribution to $S$ somewhat
\cite{swalk}, but one already knows that in QCD-like theories, the full
nonperturbative contribution to $S$ is larger than the perturbative estimate by
approximately a factor of 2 \cite{pt}, so it could be challenging to try to
reduce this sufficiently, even in a walking TC theory. The value $S \simeq 0.2$
is larger than the region of $S$ values (forming a tilted elliptical region in
an $S$-$T$ plot \cite{sth}) favored by experiment.

Therefore, to minimize the $\psi$ contributions to $S$, one would want to have
the following mass orderings: 
\beq
m_{\psi_1} < m_{\psi_2} \quad {\rm for \ SMY \ case}
\label{psimassordering_smy}
\eeq
and
\beq
m_{\psi_1} > m_{\psi_2} \quad {\rm for \ RSMY \ case} \ . 
\label{psimassordering_rsmy}
\eeq
In both cases, the second term, $-2Y_{\psi_L}\ln(m_{\psi_1}/m_{\psi_2})$, in
the square brackets in Eq. (\ref{spsi_approx}) is negative and helps to reduce
the contribution to $S$ from the first term.  Note that in both of these cases,
the mass orderings that minimize the contribution to $S$ are such that the 
neutral member of the $\psi$ doublet is lighter than the charged member. 
It will be useful to consider two illustrative sets of mass values, 
\beq
{\rm SMY}: \quad m_{\psi_1} = 120 \ {\rm GeV} \ , 
           \quad m_{\psi_2} = 160 \ {\rm GeV} 
\label{mpsi_illustrative_smy}
\eeq
and
\beq
{\rm RSMY}: \quad m_{\psi_1} = 160 \ {\rm GeV} \ , 
            \quad m_{\psi_2} = 120 \ {\rm GeV} 
\label{mpsi_illustrative_rsmy}
\eeq
as well as a continuous variation of the heavier $\psi$ in each case, with the
lighter $\psi$ fixed at the value of 120 GeV. The amount by which
$m_{\psi_2}/m_{\psi_1}$ can exceed unity in the SMY case, or
$m_{\psi_1}/m_{\psi_2}$ can exceed unity in the RSMY case, is constrained in at
least two ways.  First, it is a challenge in this model for the ETC interaction
to produce such large masses for both $\psi_1$ and $\psi_2$, and this challenge
is especially severe for the heavier of these, in the two respective cases.
Given the experimental lower limit on the lighter of the two, $m_{\psi_\ell}$,
the more one tries to increase the ratio $m_{\psi_h}/m_{\psi_\ell}$, the more
of a problem it is to achieve this with credible ETC interactions.  Second, the
larger the ratio of, and the splitting between, the masses of the heavier and
lighter of the $\psi$s, the greater is the violation of custodial symmetry and
the larger is the contribution to the $\rho$ parameter. This is given by
\beq
\Delta \rho = \frac{G_F}{8 \pi^2 \sqrt{2}} \, f(m_{\psi_1}^2,m_{\psi_2}^2) = 
\frac{f(m_{\psi_1}^2,m_{\psi_2}^2)}{16 \pi^2 v^2}  \ , 
\label{delta_rho}
\eeq
where $v=2m_W/g=246$ GeV and 
\beq
f(x,y) = x+y - \frac{2xy}{x-y} \, \ln \Big ( \frac{x}{y} \Big ) \ . 
\label{fxy}
\eeq
As is evident from Eq. (\ref{fxy}), $f(x,y)=f(y,x)$ and $f(x,x)=0$.  The
corresponding contribution to $T$ is $\Delta T = \alpha_{em}(m_Z)^{-1} \Delta
\rho$. By convention, $T$ is defined with the SM contribution from the $t$
quark removed.  Any new contribution is restricted experimentally to lie within
the tilted elliptical region in the $S$, $T$ plane \cite{sth}. With the
illustrative mass values (\ref{mpsi_illustrative_smy}) and
(\ref{mpsi_illustrative_rsmy} for the SMY and RSMY hypercharge assignments,
respectively, we find that the additional contribution from the $\psi$ fermions
to $S$ is 0.022.  For comparison, if $\psi_1$ and $\psi_2$ had degenerate
masses equal to the smaller value, $m_{\psi_1}=m_{\psi_2}=120$ GeV, then this
contribution would be 0.056, while if they had degenerate masses equal to the
larger value, $m_{\psi_1}=m_{\psi_2}=160$ GeV, then this contribution would be
0.055, so there is a significant reduction in $S$ due to the nondegeneracy in
masses of $\psi_1$ and $\psi_2$. Summing this with the technifermion
contribution, we have, for the given hypercharge assignment and these
illustrative sets of $\psi_j$ values for the SMY and RSMY cases, the total
perturbative estimate that the new $G_{EW}$-nonsinglet fermions in this model
contribute $(\Delta S)_{pert.} \simeq 0.18$. This is on the high side of the
values preferred by current global precision electroweak fits, but appears to
be admissible.  For the same $\psi_j$ masses, we find that the new contribution
to $T$ is 0.03, which is easily small enough to be allowed by experimental
constraints.  Thus, the main restriction on how large the ratio
$m_{\psi_h}/m_{\psi_\ell}$ can be comes more from the difficulty of producing
such a heavy $\psi_h$ from reasonable ETC interactions than from the $\rho$
($T$) parameter.

\begin{figure}
  \begin{center}
    \includegraphics[height=6cm]{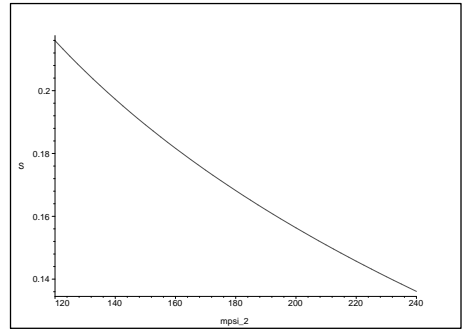}
  \end{center}
\caption{\footnotesize{
Contribution to $S$ from technifermions and the $\psi$ fermions, as a
  function of $m_{\psi_2}$ in units of GeV, with $m_{\psi_1}$ fixed at the 
illustrative value $m_{\psi_1}=120$ GeV, for the SMY hypercharge assignments 
of Eq. (\ref{smy}). The same figure applies to the RSMY hypercharge
assignments with $m_{\psi_1}$ and $m_{\psi_2}$ interchanged.}}
\label{stc}
\end{figure}

\begin{figure}
  \begin{center}
    \includegraphics[height=6cm]{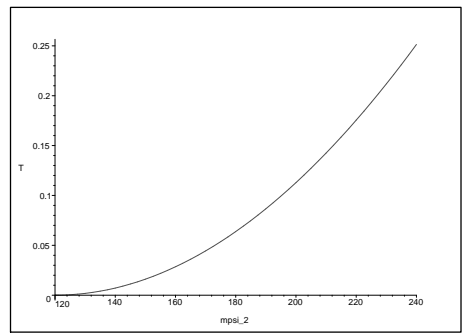}
  \end{center}
\caption{\footnotesize{ Contribution to $T$ from the $\psi$ fermions, as a
function of $m_{\psi_2}$ in units of GeV, with $m_{\psi_1}$ fixed at the
illustrative value $m_{\psi_1}=120$ GeV. The contribution to $T$ is invariant
under the interchange of $m_{\psi_1}$ and $m_{\psi_2}$.}}
\label{tpsi}
\end{figure}

Generalizing this analysis, in Fig. \ref{stc}, we show the total estimate for
$\Delta S$ from the technifermions and the $\psi$ fermions, as a function of
$m_{\psi_2} > m_{\psi_1}$ for the SMY case, with $m_{\psi_1}$ fixed at the
value of 120 GeV. In Fig. \ref{tpsi} we plot the corresponding contribution
from the nondegenerate $\psi_1$ and $\psi_2$ fermions to $T$.  The same figures
apply for the RSMY case with $m_{\psi_1}$ and $m_{\psi_2}$ interchanged. If we
restrict $\Delta T$ to be less than, say, 0.1, then this restricts the heavier
of the $\psi_j$s, i.e., $m_{\psi_2}$ for SMY and $m_{\psi_1}$ for RSMY, to be
less than approximately 195 GeV.

With the mass ordering (\ref{psimassordering_smy}) for the SMY hypercharge
case, the $(\psi_2)^-$ will decay to $(\psi_1)^0$ via a charged-current weak
transition, via a virtual $W^-$ which could produce $\ell^- \bar\nu_\ell$ with
$\ell=e, \ \mu, \tau$, as well as $d \bar u$ and $s \bar c$. Similarly, with
the mass ordering (\ref{psimassordering_rsmy}) for the RSMY hypercharge case,
the $(\psi_1)^+$ will decay to $(\psi_2)^0$, producing $\ell^+ \nu_\ell$, $u
\bar d$, and $c \bar s$ final states.  Treating these cases together, it
follows that the inclusive weak decay rate of the $\psi_h$ due to these
charged-current decays would be $\Gamma(\psi_h) = (3+2N_c)\Gamma_\mu$, up to
phase space factors reflecting the substantial mass of $\psi_\ell$ relative to
$\psi_h$. If we neglect mixing effects, then for the illustrative values
(\ref{mpsi_illustrative_smy}), using a calculation of the phase space
suppression factor, of $3 \times 10^{-3}$, from Ref. \cite{leptondecay}, we
obtain the rough estimate
\beqs
\tau_{\psi_h} & \simeq & \frac{1}{9} \times (3 \times 10^2) 
\Big ( \frac{m_\mu}{m_{\psi_2}} \Big )^5 \tau_\mu \cr\cr
& \simeq & (10^{-20} \ {\rm sec} ) \, 
\Big ( \frac{160 \ {\rm GeV}}{m_{\psi_h}}  \Big )^5 \ . 
\label{taupsi2}
\eeqs
There will also be mixing effects that would allow $\psi_h$ to undergo
charged-current decays without such phase space suppression, but this is
already a very short lifetime.  We note that, owing to the fact that the
leptonic weak neutral current contains nondiagonal terms \cite{ls}, the
$\psi_h$ could also decay via neutral-current reactions, which would reduce its
lifetime.

\section{Extended Technicolor Theories Containing 1DTC Sectors}

\subsection{General} 

So far we have studied technicolor models which have a minimal, single
SU(2)$_L$ doublet of left-handed technifermions, together with corresponding
right-handed technifermions (all of which are color-singlets), as in
Eqs. (\ref{fl}) and (\ref{fr}), focusing on the case of $G_{TC}={\rm
SU}(3)_{TC}$ with technifermions in the fundamental representation.  We have
analyzed some phenomenological restrictions on this model and have noted the
optional addition of a set of $G_{SM}$-singlet technifermions to get walking
behavior. We now proceed to analyze properties of extended technicolor theories
that contain these technicolor sectors.  We use the term ``ultraviolet
extension'' to refer to these, rather than the more ambitious term
``ultraviolet completion'', because additional ingredients would be needed to
account fully for the precise values of the fermion masses and mixings, etc.

The basic purpose of extended technicolor is to communicate the electroweak
symmetry breaking in the technifermion sector to the SM fermions, which are
technisinglets, and thereby to give them masses.  As was noted in the
introduction, because some ETC gauge bosons transform as fundamental
representations of SU(3)$_c$, it follows that commutators of the corresponding
generators of $G_{ETC}$ with their hermitian conjugates yield generators that
transform according to the singlet and adjoint representation of SU(3)$_c$,
which implies the structural property (\ref{getcform}).  With $G_{gen.}={\rm
SU}(3)_{gen.}$ and $G_{TC} = {\rm SU}(3)_{TC}$, Eq. (\ref{getcform}) takes the
form, for non-Abelian factor groups, 
\beq
G_{ETC} \supset {\rm SU}(3)_c \otimes {\rm SU}(3)_{gen.} \otimes 
{\rm SU}(3)_{TC} \ . 
\label{getcformsu3}
\eeq
In 1DTC models the ETC gauge bosons also carry weak hypercharge, $Y$.  The
structure of the ETC theory depends on whether or not one includes the
additional $G_{SM}$-singlet fermions.  To have a compact notation to refer to
these two types of 1DTC theories, we introduce the abbreviations 1DTCM and
1DTCA for the \underline{m}inimal 1DTC model and the 1DTC model
\underline{a}ugmented with the the additional $G_{SM}$-singlet technifermions.
Although the 1DTCM model, without additional ingredients, does not exhibit
walking behavior, it serves as a useful contrast to the 1DTCA model as regards
respective ETC theories.  The structural formula, Eq. (\ref{getcformsu3}) holds
for both 1DTCM and 1DTCA models; however, as we will show, $G_{ETC}$ also
includes SU(2)$_L$ as a subgroup in the case of the 1DTCA model.

\subsection{ETC Ultraviolet Extension of a 1DTCM Model} 

Although the exchanges of ETC gauge bosons that produce the masses of the SM
fermions and of the $\psi$ fermions involve strong coupling and nonperturbative
physics, the quantum numbers carried by the ETC gauge bosons can be determined
by an analysis of the basic perturbative vertices. For the 1DTCM model, one can
group the various types of fermions of a given chirality $\chi=L,R$ in the two
sets
\beq
[ \{ u^{ai} \}, \   \{ \nu^i \}, \   f^\tau_1, \  \psi_1 ]_\chi
\label{upperset}
\eeq
and
\beq
[ \{ d^{ai} \}, \   \{ \ell^i \}, \   f^\tau_2, \  \psi_2 ]_\chi \ , 
\label{lowerset}
\eeq
where $a$, $i$, and $\tau$ are color, generation, and technicolor gauge
indices, and $\{u^{ai} \}$ and $\{d^{ai} \}$ denote the respective sets of
$Q=2/3$ and $Q=-1/3$ quarks, each with $N_{gen.}N_c=9$ members, $\{\nu^i \}$
and $\{ \ell^i \}$ denote the corresponding sets of $N_{gen.}=3$ neutrinos and
leptons, and the other fermions were given in Eqs. (\ref{fl}), (\ref{fr}), and
(\ref{psi}). Some ETC-mediated transitions operate within each set. Among these
are, first, vectorial transitions of the form $q^{ai} \to q^{bi}$, where $q=u$
or $q=d$, mediated by the color gluons in the SU(3)$_c$ subgroup of
Eq. (\ref{getcform}).  Secondly, there are vectorial technicolor transitions
$f^\tau_j \to f^{\tau'}_j$, $j=1,2$, mediated by the technigluons of the
$G_{TC}$ subgroup in Eq. (\ref{getcform}).  Third, there are transitions
involving generational indices, $q^{ai} \to q^{aj}$, where $q=u$ or $q=d$,
$\nu^i \to \nu^j$, and $\ell^i \to \ell^j$, involving ETC gauge bosons in the
SU(3)$_{gen.}$ subgroup in Eq. (\ref{getcform}). Then there are the ${4 \choose
2}=6$ types of ETC-mediated transitions between each group of fermions in the
set (\ref{upperset}) and, separately, in the set (\ref{lowerset}).  Of these
six types of transitions, three enable the SM fermions and the $\psi$ fermions
to make transitions to technifermions and hence pick up masses.  The other
three involve transitions between the subsets of fermions $\{u^{ai}\}$,
$\{\nu^j\}$, and $\psi_1$ on the one hand, and, separately, between
$\{d^{ai}\}$, $\{\ell^j\}$, and $\psi_2$. Moreover, various
commutators of the nondiagonal ETC generators corresponding to these gauge
bosons and their hermitian conjugates produce diagonal generators in Cartan
subalgebras of $G_{ETC}$.

An important property of the 1DTCM model is that there is a 1--1 correspondence
between the fermions in the set (\ref{upperset}) and in the set
(\ref{lowerset}).  This reflects a kind of left-right extension of the basic
SU(2)$_L$ symmetry according to which the upper member of an SU(2)$_L$ doublet
can be transformed into the lower member of the same doublet.  In particular,
this means that all ETC-mediated transitions occur between chiral fermions 
that transform in the same way under SU(2)$_L$ (as doublets for all 
left-handed fermions, and singlets for all right-handed fermions).  Therefore,
in the 1DTCM model, all ETC gauge bosons are SU(2)$_L$ singlets, and 
\beq
[G_{ETC},{\rm SU}(2)_L] = \emptyset \quad {\rm for \ 1DTCM \ model} \ . 
\label{su2weakgetc_commute_1dtcm}
\eeq
This commutativity does not hold in 1DTCA models, as will be discussed below. 

An analysis of the basic vertices and associated transitions determines the
quantum numbers of the ETC gauge bosons.  In addition to the gauge bosons in
the SU(3)$_c$, SU(3)$_{gen.}$, and SU(3)$_{TC}$ subgroups of $G_{ETC}$, we
have, for both 1DTCM and 1DTCA models, the following transitions involving
fermions, where, as before, $a$, $i$, and $\tau$ are color, generation, and
technicolor gauge indices and $\chi=L,R$: 
\beqs
& & u^{ai}_\chi \to f^\tau_{1\chi} + V^{ai}_\tau \ , \cr\cr
& & d^{ai}_\chi \to f^\tau_{2\chi} + V^{ai}_\tau, \ , 
\label{uftransition}
\eeqs
\beqs
& & \nu^i_\chi \to f^\tau_{1\chi} + V^i_\tau \ ,  \cr\cr
& & \ell^i_\chi \to f^\tau_{2\chi} + V^i_\tau \ , 
\label{nutof}
\eeqs
and
\beqs
& & \psi_{1\chi} \to f^\tau_{1\chi} + V_\tau \ ,  \cr\cr
& & \psi_{2\chi} \to f^\tau_{2\chi} + V_\tau \ . 
\label{psitof}
\eeqs
Three other types of transitions, with their associated ETC gauge bosons, are
\beqs
& & u^{ai}_\chi \to \nu^j_\chi +  V^{ai}_j \ , \cr\cr
& & d^{ai}_\chi \to \ell^j_\chi + V^{ai}_j \ , 
\label{utonu}
\eeqs
\beqs
& & u^{ai}_\chi \to \psi_{1\chi} +  V^{ai} \ , \cr\cr
& & d^{ai}_\chi \to \psi_{2\chi} +  V^{ai} \ , 
\label{utopsi}
\eeqs
and
\beqs
& & \nu^i_\chi \to \psi_{1\chi} +  V^i \ , \cr\cr
& & \ell^i_\chi \to \psi_{2\chi}+  V^i  \ . 
\label{nutopsi}
\eeqs
One can read off from these transitions the representation content of the
associated ETC gauge bosons under the product group (\ref{getcformsu3}).  We
list these in Table \ref{etcgb_1dtcm}.  Hermitian conjugates of ETC gauge
bosons corresponding to nondiagonal generators are understood; for example,
$V^\tau_{ai} = (V^{ai}_\tau)^\dagger$, etc.  ETC gauge bosons corresponding to
diagonal, Cartan generators occur in a block-diagonal manner in accordance with
the subgroup structure of Eq. (\ref{getcformsu3}).  Since the ETC gauge bosons
are SU(2)$_L$-singlets for a 1DTCM model, it follows that their electric
charges are given by $Q_V=Y_V/2$.  

As for one-family TC/ETC models, in order for ETC interactions to account for
the generational hierarchy in SM fermion masses, the SU(3)$_{gen.}$ part of the
ETC gauge symmetry should break sequentially at scales $\Lambda_i$, $i=1,2,3$,
where $i$ is a generation index. Typical values of these scales for the
one-family TC/ETC models of Refs. \cite{nt}-\cite{kt} were listed in
Eq. (\ref{lametc}), and roughly similar values would apply here.  At each stage
of this sequential generational ETC symmetry breaking, the ETC gauge bosons
corresponding to generators in the coset space gain masses of order the
respective breaking scale. Thus, the ETC gauge bosons containing a generational
index $i$ gain masses of order $\Lambda_i$, and the ETC gauge bosons that
contain two generational indices, such as $V^i_j$, gain masses of order
$\Lambda_k$, where $k={\rm min}(i,j)$. Thus, for example, $V^1_2$ and $V^1_3$
would gain masses $\sim \Lambda_1$, etc.  Because the $\psi$ fermions must gain
masses greater than about 100 GeV, the $V_\tau$ ETC gauge bosons involved in
the transitions connecting these $\psi$'s with the technifermions must gain
masses of order the lowest ETC symmetry-breaking scale, $\Lambda_3$.  These
properties are indicated in Table \ref{etcgb_1dtcm}.  As with fermions, the
actual mass eigenstates of the vector bosons resulting from ETC symmetry
breaking would involve linear combinations of the ETC group eigenstates, in
accordance with the symmetries that are operative at the given mass
scale. (This is the ETC analogue of the mixing of the electroweak gauge bosons
of SU(2)$_L$ and U(1)$_Y$ to form the physical vector boson mass eigenstates
$W$ and $Z$ in the process of EWSB.)

We come next to the choice of a possible ETC group, $G_{ETC}$, for this 1DTCM
model.  Here the ETC group is considerably more complicated than was the case
for the one-family TC/ETC models, where one had the simplifying commutativity
property $[G_{ETC}, G_{SM}]=\emptyset$ and the relation (\ref{netc}).  One way
to embed the SU(3)$_{gen.}$ and SU(3)$_{TC}$ groups in an ETC group is to
choose the latter to be ${\rm SU}(N_{ETC})_{ETC}$ with
\beq
N_{ETC} = N_{gen.}(N_c+1)+N_{TC}+1 = 12+3+1=16 \ , 
\label{netcsu3}
\eeq
with the left-handed and right-handed chiral fermions assigned to the
vectorlike representations of SU(16)$_{ETC}$, 
\beq
{\cal F}_\chi = \left( \begin{array}{cccc}
      \{ u^{ai} \} & \{ \nu^i \}  & f^\tau_1 & \psi_1  \\
      \{ d^{ai} \} & \{ \ell^i \} & f^\tau_2 & \psi_2 \end{array}\right )_\chi
\ , \quad \chi=L,R \ . 
\label{fcal}
\eeq
This construction is somewhat analogous to the SU(14)$_{ETC}$
model of Ref. \cite{ae}, with the difference that here we use an SU(3)$_{TC}$
group rather than an SU(2)$_{TC}$ group for the reasons discussed above, which
also necessitated the inclusion of the $\psi$ fermions \cite{evanssannino}.  We
also note the toy ETC model in \cite{evanssannino} that focuses on the third
generation and can account for the $t$ quark mass.

Clearly, this SU(16)$_{ETC}$ group is a much more complicated ETC group than
the SU(5)$_{ETC}$ group for the one-family TC/ETC theory analyzed in detail in
Refs. \cite{at94}-\cite{kt}. To proceed, one would choose an appropriate set of
additional ETC-nonsinglet fermions that would render the full ETC theory a
chiral gauge theory.  A necessary property of this set of additional ETC
fermions would be that the ETC theory would be asymptotically free and, as the
reference energy scale $\mu$ decreased from large values, some of them would
form bilinear condensates in such a manner as to produce a sequential breaking
of the ETC gauge symmetry down to the residual exact SU(3)$_{TC}$ subgroup.  To
account for the mass hierarchy of the three SM fermion generations, the ETC
symmetry would break in a sequence of three scales, $\Lambda_{ETC,i}$,
$i=1,2,3$. Presumably, the breaking scales would be roughly comparable to those
of Eq. (\ref{lametc}).  Since a formula similar to Eq. (\ref{mfi}) also applies
to the mass generation for the $\psi_1$ and $\psi_2$ fermions, and since these
have to have quite large masses, it would be necessary that the $V_\tau$ ETC
gauge bosons gain masses at approximately the $\Lambda_{ETC,3}$ scale, as
indicated in Table \ref{etcgb_1dtcm}.  Some ingredients for the requisite ETC
gauge symmetry breaking could be adopted from the previous studies of
one-family models, as well as generalizations thereof \cite{genisb}.  As in the
models analyzed in Ref. \cite{lrs}, it would be necessary to break the
left-right symmetry of the representations in Eq. (\ref{fcal}) so as to avoid
conflict with experimental upper limits on right-handed charged currents.  As
was demonstrated in \cite{lrs}, this can also produce the chirally
non-symmetric weak hypercharge assignments for SM fermions.  It would also be
necessary to address all of the usual issues with ETC models, including
producing large enough SM fermion masses while respecting constraints from
flavor-changing neutral current processes, generating the large mass splitting
between the $t$ and $b$ quarks, producing very small nonzero neutrino masses,
designing the additional ETC-nonsinglet fermion sector in such a manner that
the desired sequential breaking pattern and associated condensate formation is
plausible, within the context of the most attractive channel formalism, etc.

\subsection{ETC Ultraviolet Extension of a 1DTCA Model}

In a 1DTCA model, the chiral fermions can be grouped into the sets
(\ref{upperset}) and (\ref{lowerset}) together with the set of $G_{SM}$-singlet
technifermions, which we shall label $s^\tau_{p,\chi}$, where $\chi=L,R$ and
$p$ is a copy (flavor) index.  Hence, in a 1DTCA model, the ETC-mediated
transitions between a fermion in the set (\ref{upperset}) with $\chi=L$ and
$s^\tau_{p,L}$ or between a fermion in the set (\ref{lowerset}) with $\chi=L$
and $s^\tau_{p,L}$ involve the emission of an ETC gauge boson that transforms
as the fundamental (doublet) representation of SU(2)$_L$. The commutators of
the generators corresponding to these SU(2)$_L$-doublet ETC gauge bosons and
their hermitian conjugates produce the singlet and adjoint representation of
SU(2)$_L$.  Hence, the gauge group of the ETC ultraviolet extension of a 1DTCA
model is larger than that for a 1DTCM model.  In particular, this means that
Eq. (\ref{getcform}) is expanded to
\beqs
G_{ETC} & \supset & {\rm SU}(3)_c \otimes {\rm SU}(2)_L \otimes G_{gen.} 
\otimes G_{TC} \cr\cr
& & {\rm for \ a \ 1DTCA \ model} \ , 
\label{getcform_1dtca}
\eeqs
and, in contrast to the commutativity property 
(\ref{su2weakgetc_commute_1dtcm}) for a 1DTCM model, we have
\beq
[G_{ETC},{\rm SU}(2)_L] \ne \emptyset \quad {\rm for \ a \ 1DTCA \ model} \ . 
\label{(su2weakgetc_noncommute_1dtca}
\eeq
The electric charge of an ETC gauge boson in a 1DTCA model is given by the full
formula $Q_V = T_{3,V} + (Y_V/2)$.  

The quantum numbers of the ETC gauge bosons in a 1DTCA model can be worked out
in a manner similar to those of a 1DTCM model.  They include the gauge bosons
in the 1DTCM model, as summarized in Table \ref{etcgb_1dtcm}, together with
others, generically denoted as $X$-type gauge bosons, that are involved in
transitions of fermions in the sets (\ref{upperset}) and (\ref{lowerset}) to
the $s^\tau_{p,\chi}$ fermions.  For example, the transition
\beq
Q_L^{ai \alpha} \to s^\tau_{p,L} + X^{ai \alpha}_\tau  \ , 
\label{qltos}
\eeq
where $\alpha$ is an SU(2)$_L$ gauge index, involves the emission of an ETC
gauge boson $X^{ai \alpha}$ that transforms as the fundamental representation,
$\fund$, under SU(3)$_c$, $\fund$ under SU(3)$_{gen.}$, $\overline{\fund}$
under SU(3)$_{TC}$, and $\fund$ under SU(2)$_L$, with $Y=1/3$.  The quantum
numbers of the other $X$-type ETC gauge bosons in a 1DTCA model can be worked
out in a similar manner.  Evidently, the ETC ultraviolet extension of a 1DTCA
model is more complicated than the SU(16)$_{ETC}$ extension of the 1DTCM model.

\section{Conclusions}

In this paper we have investigated some TC/ETC models with color-singlet
technifermions and a single SU(2)$_L$ doublet of technifermions.  We have
considered two types of models, with and without additional $G_{SM}$-singlet
technifermions.  We have analyzed a number of constraints on these models,
including constraints on hypercharge assignments for the technifermions and for
the associated color-singlet, technisinglet fermions $\psi$ arising from the
necessity to avoid exotically charged Coulombic bound states, on which there
are very stringent experimental upper limits.  We have also determined some
properties of ETC ultraviolet extensions of these technicolor models. The
results are of use for further studies of theories with dynamical electroweak
symmetry breaking.  Data that are forthcoming from the LHC will soon elucidate
whether electroweak symmetry breaking is, indeed, dynamical. 

\begin{acknowledgments}

This research was partially supported by the grant NSF-PHY-09-69739. 

\end{acknowledgments}

\newpage

\begin{table}
\caption{\footnotesize{Properties of ETC gauge bosons (g.b.) in 1DTCM
models. Here $\{a,b\}$, $\{i,j\}$; and $\{\tau,\tau'\}$ are SU(3)$_c$ color,
SU(3)$_{gen.}$, and SU(3)$_{TC}$ gauge indices, respectively, and $A$ denotes
adjoint representation. In the comments column, seq. bk. refers to the fact
that these gauge bosons have masses corresponding to the sequential breaking of
the SU(3)$_{gen.}$ gauge symmetry at the scales $\Lambda_i$, $i=1,2,3$. The
$Y_{F_L}$ values for the SMY and RSMY cases are given in Eqs. (\ref{smy}) and
(\ref{rsmy}).}}
\begin{center}
\begin{tabular}{|c|c|c|c|c|c|} \hline\hline
ETC g.b. & SU(3)$_c$ & SU(3)$_{gen.}$ & SU(3)$_{TC}$ & U(1)$_Y$ & 
comments \\ \hline
$V^a_b$          &   A     &    1    &  1 &  0  & exact color sym. \\
$V^i_j$          &   1     &    A    &  1 &  0  & seq. bk. \\
$V^\tau_{\tau'}$ &   1     &    1    &  A &  0  & exact TC sym.  \\ \hline
$V^{ai}_\tau$    & $\fund$ & $\fund$ & $\overline{\fund}$ & $1/3-Y_{F_L}$ & 
seq. bk. \\
$V^i_\tau$       &   1     & $\fund$ & $\overline{\fund}$ & $-1-Y_{F_L}$  &
seq. bk. \\
$V_\tau$         &   1     &    1    & $\overline{\fund}$ & $-4Y_{F_L}$   & 
bk., $\Lambda_3$\\
\hline
$V^{ai}_j$       & $\fund$ &    A    &  1 & 4/3 &  seq. bk. \\
$V^{ai}$         & $\fund$ & $\fund$ &  1 & $1/3+3Y_{F_L}$  &  seq. bk.  \\
$V^i$            &   1     & $\fund$ &  1 & $-1+3Y_{F_L}$   &  seq. bk.  \\
\hline\hline
\end{tabular}
\end{center}
\label{etcgb_1dtcm}
\end{table}


\begin{thebibliography}{99}
\bibitem{1dtc}
%
Note that the actual number of additional SU(2)$_L$ fermion doublets is
$(N_c+1)d_{R_{TC}}=4d_{R_{TC}}$ in a 1FTC model and $d_{R_{TC}}+N_\psi$ in a
1DTC model, where $d_{R_{TC}}$ is the dimension of the technifermion
representation and $N_\psi$ denotes additional technisinglet
SU(2)$_L$ doublets.
\bibitem{notone}
%
In early work on ETC, and some recent papers, ETC effects are modelled via
effective Lagrangians involving four-fermion operators ${\cal O}_j$ linking SM
fermions and technifermions, multiplied by coefficients depending on a presumed
single ETC scale, ${\cal L}_{eff} = \sum_j (c_j/\Lambda_{ETC}^2){\cal O}_j$.
For such models to be viable, a necessary requirement is to assume a large
generational hierarchy in the dimensionless coefficients $c_j$. In the 
reasonably ultraviolet-complete models of \cite{at94}-\cite{ckm}, one replaces
this assumption with a dynamical derivation of sequential breaking of ETC in
multiple scales, as in Eq. (\ref{lametc}).
\bibitem{lrsft}
%
The list of technifermion condensates excluded by vacuum alignment arguments 
in one-family TC models is given in footnote 14 of Ref. \cite{lrs}.
\bibitem{coul}
%
We assume that the charges $q_{\psi_i}$ for $i=1,2$ and, in particular, the
charge of the stable fermion $q_{\psi_\ell}$, satisfy $q_{\psi_i}\alpha_{em} <<
1$, so that a perturbative treatment of the Coulombic bound states is possible,
as with usual matter.
\bibitem{hps}
%
H.-J. He, N. Polonsky, and S. Su, Phys. Rev. D {\bf 64}, 053004 (2001). 
The normalization of $Y$ in this paper is 
different by a factor of 2 from the normalization that we use.
\bibitem{sth}
%
In SM model fits, the position of this tilted elliptical region also depends 
on an input value for the Higgs mass $m_H$.  Since technicolor models do not 
have a fundamental Higgs field, one has to replace this input with an effective
mass parametrizing the unitarization of the longitudinal vector boson
scattering amplitudes.  This mass is nominally of order $\Lambda_{TC}$.
\end{thebibliography}
\end{document}